# Analysis of Two-State Folding Using Parabolic Approximation II: Temperature-Dependence

**AUTHOR NAME:** Robert S. Sade

**AUTHOR ADDRESS:** Vinkensteynstraat 128, 2562 TV, Den Haag, Netherlands

**AUTHOR EMAIL ADDRESS:** robert.sade@gmail.com

**AUTHOR AFFILIATION:** Independent Researcher

# ABSTRACT

Equations that govern the temperature-dependence of the rate constants, Gibbs energies, enthalpies, entropies and heat capacities of activation for folding and unfolding of spontaneously-folding fixed two-state systems have been derived using a procedure that treats the denatured and the native conformers as being confined to harmonic Gibbs energy wells. The notion that a two-state system is physically defined only for a set temperature range is introduced. The implications of this novel treatment for protein folding are discussed.

# INTRODUCTION

It was shown previously, henceforth referred to as Paper-I, that the equilibrium and kinetic behaviour of spontaneously-folding fixed two-state proteins can be analysed by a treatment that is analogous to that given by Marcus for electron transfer.[1,2] In this framework termed the parabolic approximation, the Gibbs energy functions of the denatured (DSE) and the native state (NSE) ensembles are represented by parabolas whose curvature is given by their temperature-invariant force constants, $\alpha$ and $\omega$, respectively. The temperature-invariant mean length of the reaction coordinate (RC) is given by $m_{D-N}$ and is identical to the separation between the vertices of the DSE and the NSE parabolas along the abscissa. Similarly, the position of the transition state ensemble (TSE) relative to the DSE and the NSE are given by $m_{TS-D(T)}$ and $m_{TS-N(T)}$, respectively, and are identical to the separation between the *curve-crossing* and the vertices of the DSE and the NSE parabolas along the abscissa, respectively. The Gibbs energy of unfolding at equilibrium, $\Delta G_{D-N(T)}$, is identical to the separation between the vertices of the DSE and the NSE parabolas along the ordinate. Similarly, the Gibbs activation energy for folding ($\Delta G_{TS-D(T)}$) and unfolding ($\Delta G_{TS-N(T)}$) are identical to the separation between the *curve-crossing* and the vertices of the DSE and the NSE parabolas along the ordinate, respectively (**Figure 1**). Further, it was shown that the *curve-crossing* relative to the DSE and the NSE Gibbs basins are given by

$$m_{TS-D(T)} = \frac{\omega m_{D-N} - \sqrt{\lambda\omega + \Delta G_{D-N(T)}(\omega-\alpha)}}{(\omega-\alpha)} = \frac{\omega m_{D-N} - \sqrt{\varphi}}{(\omega-\alpha)} \tag{1}$$

$$m_{TS-N(T)} = \frac{\sqrt{\lambda\omega + \Delta G_{D-N(T)}(\omega-\alpha)} - \alpha m_{D-N}}{(\omega-\alpha)} = \frac{\sqrt{\varphi} - \alpha m_{D-N}}{(\omega-\alpha)} \tag{2}$$

where the discriminant $\varphi = \lambda\omega + \Delta G_{D-N(T)}(\omega-\alpha)$, and $\lambda = \alpha(m_{D-N})^2$ is the *Marcus reorganization energy* for two-state folding, and by definition is the Gibbs energy required to compress the denatured polypeptide under folding conditions to a state whose solvent accessible surface area (SASA) is identical to that of the native folded protein but without the stabilizing native interactions. If the temperature-dependence of $\Delta G_{D-N(T)}$ and the values of $\alpha$, $\omega$, and $m_{D-N}$ are known for any two-state system at constant pressure and solvent conditions (see **Methods** in Paper-I), the position of its TSE relative to the ground states for any temperature may be readily calculated for that particular solvent. The temperature-dependence of the *curve-crossing* is central to this analysis since all relevant thermodynamic state functions can be readily derived by manipulating the same.[3] Because by postulate the force constants and $m_{D-N}$ are temperature-invariant for a fixed two-state folder at constant pressure and solvent conditions, we get from inspection of Eqs. (1) and (2) that the discriminant $\varphi$, and $\sqrt{\varphi}$ must be a maximum when $\Delta G_{D-N(T)}$ is a maximum. Because $\Delta G_{D-N(T)}$ is a maximum when $T = T_S$ ($T_S$ is the temperature at which the entropy of unfolding at equilibrium, $\Delta S_{D-N(T)}$, is zero),[4] a corollary is that $\varphi$ and $\sqrt{\varphi}$ must be a maximum when $T = T_S$; and any deviation in the temperature from $T_S$ will only lead to their decrease. Consequently, $m_{TS-D(T)}$ and $\beta_{T(fold)(T)}$ ($= m_{TS-D(T)}/m_{D-N}$) are always a minimum, and $m_{TS-N(T)}$ and $\beta_{T(unfold)(T)}$ ($= m_{TS-N(T)}/m_{D-N}$) are always a maximum at $T_S$. This gives rise to two further corollaries: Any deviation in the temperature from $T_S$ can only lead to: (*i*) an increase in $m_{TS-D(T)}$ and $\beta_{T(fold)(T)}$; and (*ii*) a decrease in $m_{TS-N(T)}$ and $\beta_{T(unfold)(T)}$. A further consequence of $m_{TS-D(T)}$ being a minimum at $T_S$ is that if for a two-state-folding primary sequence there exists a chevron with a well-defined linear folding arm at $T_S$, then $m_{TS-D(T)} > 0$ for all temperatures (**Figure 1B**). Since the *curve-crossing* is physically undefined for $\varphi < 0$ owing to there being no real roots, the maximum theoretically possible value of $m_{TS-D(T)}$ will occur when $\varphi = 0$ and is given by: $m_{TS-D(T)}\big|_{max} = \omega m_{D-N}/(\omega - \alpha) > 0$ (note that since the force constants are positive, $\omega > (\omega - \alpha) \Rightarrow m_{TS-D(T)}\big|_{max} > m_{D-N}$). Because $m_{D-N} = m_{TS-D(T)} + m_{TS-N(T)}$ for a two-state system, and $m_{D-N}$ is temperature-invariant by postulate, the theoretical minimum of $m_{TS-N(T)}$ is given by: $m_{TS-N(T)}\big|_{min} = -\alpha m_{D-N}/(\omega - \alpha)$. Importantly, since $m_{TS-N(T)}$ is a maximum and positive at $T_S$ but its minimum is negative, a consequence is that $m_{TS-N(T)} = 0$ at two unique temperatures, one in the ultralow and the other in the high temperature regime. We will discuss how this can lead to *barrierless unfolding* and *inversion of the unfolding rate constant* later. The temperature-dependent shift in the *curve-crossing* relative to the ground states along the RC is consistent with Hammond movement; and just as it is commonplace in physical organic chemistry to rationalize the physical basis of these effects using Marcus theory, we can similarly rationalize these effects in protein folding using parabolic approximation.[5-9]

The Gibbs barrier heights for folding and unfolding are given by

$$\Delta G_{\text{TS-D}(T)} = \alpha \left( m_{\text{TS-D}(T)} \right)^2 = \frac{\alpha \left( \omega m_{\text{D-N}} - \sqrt{\varphi} \right)^2}{(\omega - \alpha)^2} = \lambda \beta_{T(\text{fold})(T)}^2 \qquad (3)$$

$$\Delta G_{\text{TS-N}(T)} = \omega \left( m_{\text{TS-N}(T)} \right)^2 = \frac{\omega \left( \sqrt{\varphi} - \alpha m_{\text{D-N}} \right)^2}{(\omega - \alpha)^2} = \frac{\omega}{\alpha} \lambda \beta_{T(\text{unfold})(T)}^2 \qquad (4)$$

Because the force constants are temperature-invariant by postulate, and $m_{\text{TS-D}(T)}$ is a minimum, and $m_{\text{TS-N}(T)}$ a maximum at $T_S$, $\Delta G_{\text{TS-D}(T)}$ and $\Delta G_{\text{TS-N}(T)}$ are always a minimum and a maximum, respectively, at $T_S$. Thus, a corollary is that any deviation in temperature from $T_S$ can only lead to an increase in $\Delta G_{\text{TS-D}(T)}$ and a decrease in $\Delta G_{\text{TS-N}(T)}$. The Arrhenius expressions for the rate constants for folding ($k_{f(T)}$) and unfolding ($k_{u(T)}$), and $\Delta G_{\text{D-N}(T)}$ are given by

$$k_{f(T)} = k^0 \exp\left( -\frac{\alpha \left( \omega m_{\text{D-N}} - \sqrt{\varphi} \right)^2}{RT(\omega - \alpha)^2} \right) = k^0 \exp\left( -\frac{\lambda \beta_{T(\text{fold})(T)}^2}{RT} \right) \qquad (5)$$

$$k_{u(T)} = k^0 \exp\left( -\frac{\omega \left( \sqrt{\varphi} - \alpha m_{\text{D-N}} \right)^2}{RT(\omega - \alpha)^2} \right) = k^0 \exp\left( -\frac{\lambda \omega \beta_{T(\text{unfold})(T)}^2}{\alpha RT} \right) \qquad (6)$$

$$\Delta G_{\text{D-N}(T)} = \lambda \left( \frac{\omega}{\alpha} \beta_{T(\text{unfold})(T)}^2 - \beta_{T(\text{fold})(T)}^2 \right) = \lambda \left( \frac{\sigma_{\text{DSE}(T)}^2}{\sigma_{\text{NSE}(T)}^2} \beta_{T(\text{unfold})(T)}^2 - \beta_{T(\text{fold})(T)}^2 \right) \qquad (7)$$

where $k^0$ is the temperature-invariant prefactor with units identical to those of the rate constants (s$^{-1}$), $R$ is the gas constant, $T$ is the absolute temperature, and $\sigma_{\text{DSE}(T)}^2 = RT/2\alpha$ and $\sigma_{\text{NSE}(T)}^2 = RT/2\omega$ are the variances of the Gaussian distribution of the SASA of the conformers in the DSE and the NSE, respectively (see Paper-I). A consequence of Eq. (7) is that if for two-state systems at constant pressure and solvent conditions $\beta_{T(\text{fold})(T)} \geq 0.5$ when $T = T_S$, then it is theoretically impossible for such systems to be stable at equilibrium ($\Delta G_{\text{D-N}(T)} > 0$) unless $\sigma_{\text{NSE}(T)}^2 < \sigma_{\text{DSE}(T)}^2$.

The purpose of this article is to derive equations that will give a rigorous thermodynamic description of fixed two-state systems across a wide temperature regime. Because this article is primarily an extension of Paper-I, any critical evaluation by the reader of the interpretation and the conclusions drawn here must be done in conjunction with the framework developed in Paper-I.

In this article as well as in all subsequent papers, terms such as "favourable and unfavourable" will be used extensively. Although perhaps unnecessary, it is nevertheless mentioned that their usage is inextricably linked to the specific reaction-direction being addressed. If the reaction-direction is reversed, the magnitude of the change in the relevant state functions will be invariant, but their algebraic signs will invert leading to a change in the interpretation. For example, if the forward reaction is *endothermic* and thus *enthalpically disfavoured* at a particular temperature, it naturally implies that the reverse reaction will be *exothermic* and *enthalpically favoured*. Similarly, if $\Delta G_{D-N(T)} > 0$ at a certain temperature, then $\Delta G_{N-D(T)} < 0$; consequently, the unfolding reaction $N \rightleftharpoons D$ is *energetically disfavoured*, while the folding reaction $D \rightleftharpoons N$ is *energetically favoured* at that particular temperature. Further, the term "equilibrium stability or stability," which we will often use is synonymous with $\Delta G_{D-N(T)}$ and not $\Delta G_{N-D(T)}$. The reader will also note that in addition to the standard reference temperatures used in protein folding,[4] a few novel reference temperatures are introduced here to enable a physical description of the system. A glossary of the same is given in **Table 1**.

## THERMODYNAMIC RELATIONSHIPS

### Activation entropy for folding ($\Delta S_{TS-D(T)}$)

The activation entropy for the partial folding reaction $D \rightleftharpoons [TS]$ ($D$ denotes DSE and $[TS]$ denotes TSE) is given by the first derivative of $\Delta G_{TS-D(T)}$ with respect to temperature. Differentiating Eq. (3) gives

$$\frac{\partial \Delta G_{TS-D(T)}}{\partial T} = \frac{\partial}{\partial T} \alpha \left( m_{TS-D(T)} \right)^2 = -\Delta S_{TS-D(T)}$$

$$\Rightarrow \Delta S_{TS-D(T)} = -\alpha \frac{\partial \left( m_{TS-D(T)} \right)^2}{\partial T} = -2\alpha\, m_{TS-D(T)} \frac{\partial m_{TS-D(T)}}{\partial T} \qquad (8)$$

Note that since $m_{TS-D(T)}$ is a composite function, we use the chain rule. The partial derivatives are to indicate that the derivative is at constant pressure and solvent conditions (constant pH, ionic strength, co-solvents etc.); however, we will omit the subscripts for brevity. Substituting Eqs. (A6) and (A9) in (8) gives (see **Appendix**)

$$\Delta S_{TS-D(T)} = -\frac{\alpha\, m_{TS-D(T)} \Delta S_{D-N(T)}}{\sqrt{\varphi}} = \frac{\alpha\, m_{TS-D(T)} \Delta C_{pD-N}}{\sqrt{\varphi}} \ln\left(\frac{T_S}{T}\right) \qquad (9)$$

Although $\alpha$ and $\Delta C_{pD-N}$ are positive and temperature-invariant by postulate, since $\sqrt{\varphi}$ has no real roots for $\varphi < 0$, $m_{TS-D(T)} > 0$ no matter what the temperature, and both $\sqrt{\varphi}$ and $m_{TS-D(T)}$ vary with temperature, the temperature-dependence of $\Delta S_{TS-D(T)}$ would be complex, with its algebraic sign being determined purely by the $\ln(T_S/T)$ term. This leads to three scenarios: (*i*) $\Delta S_{TS-D(T)} > 0$ for $T < T_S$; (*ii*) $\Delta S_{TS-D(T)} < 0$ for $T > T_S$; (*iii*) $\Delta S_{TS-D(T)} = 0$ when $T = T_S$. Thus, the activation of denatured conformers to the TSE is entropically: (*i*) favourable for $T < T_S$; (*ii*) unfavourable for $T > T_S$; and (*iii*) neither favourable nor unfavourable when $T = T_S$.

## Activation entropy for unfolding ($\Delta S_{TS-N(T)}$)

The activation entropy for the partial unfolding reaction $N \rightleftharpoons [TS]$ ($N$ denotes NSE) may be similarly obtained by differentiating $\Delta G_{TS-N(T)}$ (Eq. (4)) with respect to temperature. Thus, we may write

$$\frac{\partial \Delta G_{TS-N(T)}}{\partial T} = \frac{\partial}{\partial T} \omega\left(m_{TS-N(T)}\right)^2 = -\Delta S_{TS-N(T)}$$
$$\Rightarrow \Delta S_{TS-N(T)} = -\omega \frac{\partial \left(m_{TS-N(T)}\right)^2}{\partial T} = -2\omega\, m_{TS-N(T)} \frac{\partial m_{TS-N(T)}}{\partial T} \qquad (10)$$

Substituting Eq. (A11) in (10) gives (see **Appendix**)

$$\Delta S_{TS-N(T)} = \frac{\omega\, m_{TS-N(T)} \Delta S_{D-N(T)}}{\sqrt{\varphi}} = \frac{\omega\, m_{TS-N(T)} \Delta C_{pD-N}}{\sqrt{\varphi}} \ln\left(\frac{T}{T_S}\right) \qquad (11)$$

Despite the apparent similarity between Eqs. (9) and (11), since $m_{TS-N(T)}$ unlike $m_{TS-D(T)}$ can be positive, zero, or even negative depending on the temperature, the variation in the algebraic sign of the $\Delta S_{TS-N(T)}$ function with temperature, and its physical interpretation is far more complex. Although it is readily apparent that $\Delta S_{\text{TS-N}(T)}$ must be zero when $T = T_S$, theoretically there can be two additional temperatures at which $\Delta S_{\text{TS-N}(T)}$ is zero, one in the ultralow temperature regime (designated $T_{S(\alpha)}$), and the other at high temperature (designated $T_{S(\omega)}$); and these two additional temperatures at which $\Delta S_{\text{TS-N}(T)}$ is zero occur when $m_{\text{TS-N}(T)} = 0$. Thus, the algebraic sign of the $\Delta S_{\text{TS-N}(T)}$ function across a wide temperature regime is determined by both $m_{\text{TS-N}(T)}$ and the $\ln(T/T_S)$ terms: (*i*) for $T < T_{S(\alpha)}$, both $m_{\text{TS-N}(T)}$ and $\ln(T/T_S)$ are negative, leading to $\Delta S_{\text{TS-N}(T)} > 0$; (*ii*) when $T = T_{S(\alpha)}$, $m_{\text{TS-N}(T)} = 0$, leading to $\Delta S_{\text{TS-N}(T)} = 0$; (*iii*) for $T_{S(\alpha)} < T < T_S$, $m_{\text{TS-N}(T)} > 0$ but $\ln(T/T_S) < 0$, leading to $\Delta S_{\text{TS-N}(T)} < 0$; (*iv*) when $T = T_S$, $\ln(T/T_S) = 0$, leading to $\Delta S_{\text{TS-N}(T)} = 0$; (*v*) for $T_S < T < T_{S(\omega)}$, both $m_{\text{TS-N}(T)}$ and $\ln(T/T_S)$ are positive, leading to $\Delta S_{\text{TS-N}(T)} > 0$; (*vi*) when $T = T_{S(\omega)}$, $m_{\text{TS-N}(T)} = 0$, leading to $\Delta S_{\text{TS-N}(T)} = 0$; and (*vii*) for $T > T_{S(\omega)}$, $m_{\text{TS-N}(T)} > 0$ but $\ln(T/T_S) < 0$, leading to $\Delta S_{\text{TS-N}(T)} < 0$. Consequently, we may state that the activation of native conformers to the TSE is entropically: (*i*) favourable for $T < T_{S(\alpha)}$ and $T_S < T < T_{S(\omega)}$; (*ii*) unfavourable for $T_{S(\alpha)} < T < T_S$ and $T > T_{S(\omega)}$; and (*iii*) neither favourable nor unfavourable at $T_{S(\alpha)}$, $T_S$, and $T_{S(\omega)}$. If we reverse the reaction-direction (i.e., the partial folding reaction $[TS] \rightleftharpoons N$), the algebraic sign of the $\Delta S_{\text{TS-N}(T)}$ function inverts leading to a change in the interpretation. Therefore, we may state that the flux of the conformers from the TSE to the NSE is entropically: (*i*) unfavourable for $T < T_{S(\alpha)}$ and $T_S < T < T_{S(\omega)}$ ($\Delta S_{\text{TS-N}(T)} > 0 \Rightarrow \Delta S_{\text{N-TS}(T)} < 0$); (*ii*) favourable for $T_{S(\alpha)} < T < T_S$ and $T > T_{S(\omega)}$ ($\Delta S_{\text{TS-N}(T)} < 0 \Rightarrow \Delta S_{\text{N-TS}(T)} > 0$); and (*iii*) neutral at $T_{S(\alpha)}$, $T_S$, and $T_{S(\omega)}$. Note that the term "flux" is operationally defined as the "diffusion of the conformers from one reaction state to the other on the Gibbs energy surface."

We know from the pioneering work on the temperature-dependence of protein stability that at $T_S$ we have $\Delta S_{\text{D-N}(T)} = \Delta S_{\text{TS-N}(T)} - \Delta S_{\text{TS-D}(T)} = 0$.[4,10] However, this condition *per se* does not give us any information on the absolute values of $\Delta S_{\text{TS-D}(T)}$ or $\Delta S_{\text{TS-N}(T)}$ other than tell us that they must be identical at $T_S$. Eqs. (9) and (11) are remarkable because they demonstrate that $\Delta S_{\text{TS-D}(T)}$ and $\Delta S_{\text{TS-N}(T)}$ are independently equal to zero at $T_S$. Consequently, at $T_S$ we have $S_{\text{D}(T)} = S_{\text{TS}(T)} = S_{\text{N}(T)}$. This relationship must hold true for every two-state folder when pressure and solvent are constant.

These analyses lead to two fundamentally important conclusions: First, for any two-state folder at constant pressure and solvent conditions, $\Delta G_{D-N(T)}$ will be the greatest when the activation entropies for folding and unfolding are both zero, and this occurs precisely at $T_S$. A corollary is that $\Delta G_{D-N(T)}$ is a maximum when $\Delta G_{TS-D(T)}$ and $\Delta G_{TS-N(T)}$ are both purely enthalpic; and because $\Delta G_{TS-D(T)}$ and $\Delta G_{TS-N(T)}$ are both positive at $T_S$, it implies that the activation enthalpy for folding ($\Delta H_{TS-D(T)}$) and unfolding ($\Delta H_{TS-N(T)}$) are both endothermic at $T_S$. Because by postulate $m_{D-N}$, $m_{TS-D(T)}$ and $m_{TS-N(T)}$ are true proxies for $\Delta SASA_{D-N}$, $\Delta SASA_{D-TS(T)}$ and $\Delta SASA_{TS-N(T)}$, respectively, "*equilibrium stability of a two-state system at constant pressure and solvent conditions will be the greatest when the conformers in the DSE are displaced the least from the mean of their ensemble along the SASA-RC to reach the TSE (or bury the least amount of SASA to reach the TSE), and this occurs precisely at $T_S$.*" This "*principle of least displacement*" must be valid for every two-state system. A corollary is that $\Delta G_{D-N(T)}$ will be the greatest when the *native conformers* expose the greatest amount of SASA to reach the TSE.

Second, although $\Delta S_{TS-N(T)} = 0 \Rightarrow S_{TS(T)} = S_{N(T)}$ at $T_{S(\alpha)}$, $T_S$, and $T_{S(\omega)}$, the underlying thermodynamics is fundamentally different at $T_S$ as compared to $T_{S(\alpha)}$ and $T_{S(\omega)}$. While both $\Delta G_{TS-N(T)}$ and $m_{TS-N(T)}$ are a maximum and $\Delta G_{TS-N(T)}$ is purely enthalpic at $T_S$ ($\Delta G_{TS-N(T)} = \Delta H_{TS-N(T)}$), we have $m_{TS-N(T)} = 0 \Rightarrow \Delta G_{TS-N(T)} = \omega(m_{TS-N(T)})^2 = 0 \Rightarrow \Delta H_{TS-N(T)} = 0$ at $T_{S(\alpha)}$ and $T_{S(\omega)}$. In addition, because $\Delta G_{TS-N(T)} = 0$ at $T_{S(\alpha)}$ and $T_{S(\omega)}$, the rate constant for unfolding will reach a maximum for that particular solvent at these two temperatures. To summarize, while $G_{TS(T)} \gg G_{N(T)}$ and $S_{D(T)} = S_{TS(T)} = S_{N(T)}$ at $T_S$, we have $G_{TS(T)} = G_{N(T)}$, $H_{TS(T)} = H_{N(T)}$, $S_{TS(T)} = S_{N(T)}$, and $k_{u(T)} = k^0$ at $T_{S(\alpha)}$ and $T_{S(\omega)}$. A corollary is that if two reaction-states on a protein folding pathway have identical SASA and Gibbs energy under identical environmental conditions (temperature, pressure, pH, co-solvents etc.), then their enthalpies and entropies, must also be identical. This must hold irrespective of whether or not they have identical, similar, or dissimilar molecular structures. The implications of this statement for the applicability of the Hammond postulate to protein folding will be addressed in the subsequent publication.

**Temperature limits of a two-state system**

A singularly important consequence of $\sqrt{\varphi}$ being undefined for $\varphi < 0$ has a deeper physical meaning. It implies that a barrier-limited two-state system (i.e., the notion that the conformers are confined to two intersecting harmonic Gibbs energy wells) is physically defined only for a set temperature-range given by $T_\alpha \leq T \leq T_\omega$, where $T_\alpha$ and $T_\omega$ are the ultralow and the high temperature limits, respectively. Thus, the prediction is that for $T < T_\alpha$ and $T > T_\omega$, the TSE cannot be physically defined, and all of the conformers will be confined to a single harmonic Gibbs energy well, which is the DSE (glass transition is discussed elsewhere).[11-16] A consequence is that although one could, in principle, use the Gibbs-Helmholtz equation (Eq. (A51)) and calculate equilibrium stability for any temperature above absolute zero, equilibrium stability *per se* has physical meaning only for $T_\alpha \leq T \leq T_\omega$. This is because equilibrium stability is an equilibrium manifestation of the underlying thermal-noise-driven flux of the conformers from the DSE to the NSE, and *vice versa*, *via* the TSE; consequently, if the position of the TSE along the RC becomes physically undefined owing to $\sqrt{\varphi}$ being mathematically undefined for $\varphi < 0$, $k_{f(T)}$ and $k_{u(T)}$ become physically undefined, leading to $\Delta G_{\text{D-N}(T)} = RT \ln\left(k_{f(T)}/k_{u(T)}\right)$ being physically undefined. Thus, the limit of equilibrium stability below which a two-state system becomes physically undefined is given by: $\Delta G_{\text{D-N}(T)}\big|_{T=T_\alpha, T_\omega} = -\lambda\omega/(\omega-\alpha)$ and is purely a function of the primary sequence when pressure and solvent are defined. Consequently, the physically meaningful range of equilibrium stability for a two-state system is given by: $\Delta G_{\text{D-N}(T)}\big|_{T=T_S} - \Delta G_{\text{D-N}(T)}\big|_{T=T_\alpha, T_\omega} = \Delta G_{\text{D-N}(T_S)} + \left[\lambda\omega/(\omega-\alpha)\right]$. This is akin to the stability range over which Marcus theory is physically realistic (see Kresge, 1973, page 494).[17] These arguments apply to equilibrium enthalpies and entropies as well, i.e., although the values $\Delta S_{\text{D-N}(T)}$ and $\Delta H_{\text{D-N}(T)}$ can be calculated for any temperature above absolute zero using Eqs. (A8) and (A24), respectively, they do not have any physical meaning for $T < T_\alpha$ and $T > T_\omega$. Now, from the view point of the physics of phase transitions, $T_\alpha \leq T \leq T_\omega$ denotes the *coexistence temperature-range* where the DSE and the NSE, which are in a dynamic equilibrium, will coexist as two distinct phases; and for $T < T_\alpha$ and $T > T_\omega$ there will be a single phase, which is the DSE, with $T_\alpha$ and $T_\omega$ being the limiting temperatures for coexistence (or phase boundary temperatures from the view point of the DSE), and the protein will cease to function.[18] A consequence is that as long as the covalent structure of spontaneously-folding primary sequences are not altered on exposure to $T < T_\alpha$ (as in the case of glacial periods in an ice age) and $T > T_\omega$ (as in the case of intense forest fires), their behaviour will be identical to that of untreated proteins when the temperature returns to $T_\alpha \leq T \leq T_\omega$. Further discussion on the temperature-range over which two-state systems are physically defined and how this range can be modulated by living systems to cope with a wide variety of environments, the parallels between two-state proteins and Boolean circuits, and why higher intelligence may not be possible without temperature control and biological membranes is beyond the scope of this article and will be explored in subsequent publications.[19-23]

**Activation enthalpy for folding**

The activation enthalpy for the partial folding reaction $D \rightleftharpoons [TS]$ is given by the Gibbs equation

$$\Delta H_{\text{TS-D}(T)} = \Delta G_{\text{TS-D}(T)} + T\Delta S_{\text{TS-D}(T)} \tag{12}$$

Substituting Eqs. (3) and (9) in (12) and recasting gives

$$\Delta H_{\text{TS-D}(T)} = \alpha \left(m_{\text{TS-D}(T)}\right)^2 \left[1 + \frac{T\Delta C_{p\text{D-N}}}{m_{\text{TS-D}(T)}\sqrt{\varphi}} \ln\left(\frac{T_S}{T}\right)\right] \tag{13}$$

Inspection of Eq. (13) shows that for $T < T_S$, $\Delta H_{\text{TS-D}(T)} > 0$, but decreases with a rise in temperature. When $T = T_S$, Eq. (13) reduces to $\Delta H_{\text{TS-D}(T)} = \alpha\left(m_{\text{TS-D}(T)}\right)^2 = \Delta G_{\text{TS-D}(T)} > 0$, i.e., the Gibbs barrier to folding is purely enthalpic at $T_S$, with $k_{f(T)}$ being given by

$$k_{f(T)}\Big|_{T=T_S} = k^0 \exp\left(-\frac{\Delta G_{\text{TS-D}(T)}}{RT}\right)\Bigg|_{T=T_S} = k^0 \exp\left(-\frac{\Delta H_{\text{TS-D}(T)}}{RT}\right)\Bigg|_{T=T_S} \tag{14}$$

Because $\ln(T_S/T) < 0$ for $T > T_S$ and becomes increasingly negative with a rise in temperature, as long as $1 + \left(T\Delta C_{p\text{D-N}}\ln(T_S/T)/m_{\text{TS-D}(T)}\sqrt{\varphi}\right) > 0$, $\Delta H_{\text{TS-D}(T)}$ will continue to be positive. Precisely when $T \ln(T/T_S) = m_{\text{TS-D}(T)} \sqrt{\varphi} / \Delta C_{p\text{D-N}}$, $\Delta H_{\text{TS-D}(T)} = 0$; and if we designate this temperature as $T_{H(\text{TS-D})}$, then $k_{f(T)}$ is given by

$$k_{f(T)}\Big|_{T=T_{H(\text{TS-D})}} = k^0 \exp\left(-\frac{\Delta G_{\text{TS-D}(T)}}{RT}\right)\Bigg|_{T=T_{H(\text{TS-D})}} = k^0 \exp\left(\frac{\Delta S_{\text{TS-D}(T)}}{R}\right)\Bigg|_{T=T_{H(\text{TS-D})}} \qquad (15)$$

Substituting Eq. (9) in (15) gives

$$k_{f(T)}\Big|_{T=T_{H(\text{TS-D})}} = k^0 \exp\left(\frac{\alpha \, m_{\text{TS-D}(T)} \Delta C_{p\text{D-N}}}{R\sqrt{\varphi}} \ln\left(\frac{T_S}{T}\right)\right)\Bigg|_{T=T_{H(\text{TS-D})}} \qquad (16)$$

As the temperature increases beyond $T_{H(\text{TS-D})}$, $1 + \left(T \Delta C_{p\text{D-N}} \ln(T_S/T) / m_{\text{TS-D}(T)} \sqrt{\varphi}\right) < 0$, leading to $\Delta H_{\text{TS-D}(T)} < 0$ for $T > T_{H(\text{TS-D})}$. Although we have discussed the temperature-dependence of the algebraic sign of the $\Delta H_{\text{TS-D}(T)}$ function from the view point of the sum of the terms inside the square bracket on the RHS of Eq. (13), there is another equally valid set of logical arguments that enables us to arrive at the same. Since $\alpha > 0$, we have $\Delta G_{\text{TS-D}(T)} > 0$ for all temperatures (except for the special case of *barrierless folding*, see below). Because $\Delta S_{\text{TS-D}(T)} > 0$ for $T < T_S$, (Eq. (9)) the only way the condition that $\Delta G_{\text{TS-D}(T)} = \Delta H_{\text{TS-D}(T)} - T \Delta S_{\text{TS-D}(T)} > 0$ is satisfied is if $\Delta H_{\text{TS-D}(T)} > 0$ or endothermic. In addition, when $T = T_S$, $\Delta S_{\text{TS-D}(T)} = 0$ (Eq. (9)), and the positive Gibbs activation energy for folding is entirely due to endothermic activation enthalpy for folding, i.e., $\Delta G_{\text{TS-D}(T)} = \Delta H_{\text{TS-D}(T)} > 0$. Thus, we may state that $\Delta H_{\text{TS-D}(T)} > 0$ or endothermic for $T \leq T_S$. At $T_{H(\text{TS-D})}$ we have $\Delta G_{\text{TS-D}(T)} = -T \Delta S_{\text{TS-D}(T)} > 0$ which is satisfied if and only if $\Delta S_{\text{TS-D}(T)} < 0$. Because $\Delta S_{\text{TS-D}(T)} < 0$ only when $T > T_S$ (Eq. (9)), we have the second logical condition: $T_{H(\text{TS-D})} > T_S$. Now since $\Delta G_{\text{TS-D}(T)}$ has only one extremum, which is a minimum at $T_S$ along the temperature axis ($\partial \Delta G_{\text{TS-D}(T)} / \partial T = -\Delta S_{\text{TS-D}(T)} = 0$), there will also be one extremum which is a minimum for $\Delta G_{\text{TS-D}(T)} / T$ across the temperature axis (i.e., the first derivative of the Massieu-Planck activation potential for folding is zero, or $\partial \left(\Delta G_{\text{TS-D}(T)} / T\right) / \partial T = -\Delta H_{\text{TS-D}(T)} / T^2 = 0$; see Schellman, 1997, on the use of Massieu-Planck functions in protein thermodynamics).[24] In other words, there exists only one temperature, $T_{H(\text{TS-D})}$, at which $\Delta H_{\text{TS-D}(T)} = 0$. Thus, a corollary is that for two-state folders at constant pressure and solvent conditions, "$k_{f(T)}$ is a maximum when the Gibbs barrier to folding is purely entropic." Thus, we may conclude that for a two-state system at constant pressure and solvent conditions: $\Delta H_{TS-D(T)} > 0$ at very low temperature, decreases in magnitude with a rise in temperature, and eventually reaches zero at $T_{H(TS-D)}$; and any further increase in temperature beyond $T_{H(TS-D)}$ causes $\Delta H_{TS-D(T)}$ to become negative or exothermic. To summarize, we have three important scenarios: (*i*) $\Delta H_{TS-D(T)} > 0$ for $T_\alpha \leq T < T_{H(TS-D)}$; (*ii*) $\Delta H_{TS-D(T)} < 0$ for $T_{H(TS-D)} < T \leq T_\omega$; and (*iii*) $\Delta H_{TS-D(T)} = 0$ for $T = T_{H(TS-D)}$. Thus, the activation of the denatured conformers to the TSE is enthalpically: (*i*) unfavourable for $T_\alpha \leq T < T_{H(TS-D)}$; (*ii*) favourable for $T_{H(TS-D)} < T \leq T_\omega$; and (*iii*) neutral when $T = T_{H(TS-D)}$.

**Activation enthalpy for unfolding**

The activation enthalpy for the partial unfolding reaction $N \rightleftharpoons [TS]$ is given by

$$\Delta H_{TS-N(T)} = \Delta G_{TS-N(T)} + T\Delta S_{TS-N(T)} \tag{17}$$

Substituting Eqs. (4) and (11) in (17) and recasting gives

$$\Delta H_{TS-N(T)} = \omega \left(m_{TS-N(T)}\right)^2 \left[1 + \frac{T\Delta C_{pD-N}}{m_{TS-N(T)}\sqrt{\varphi}}\ln\left(\frac{T}{T_S}\right)\right] \tag{18}$$

Unlike the $\Delta H_{TS-D(T)}$ function which is positive for $T_\alpha \leq T < T_{H(TS-D)}$, negative for $T_{H(TS-D)} < T \leq T_\omega$, and importantly, changes its algebraic sign only once across the entire temperature range over which a two-state system is physically defined, the behaviour of $\Delta H_{TS-N(T)}$ function is far more complex; and just as the $\Delta S_{TS-N(T)}$ function can become zero at three distinct temperatures, so too can the $\Delta H_{TS-N(T)}$ function. Starting from the lowest temperature ($T_\alpha$) at which a two-state system is physically defined, for $T_\alpha \leq T < T_{S(\alpha)}$, both $m_{TS-N(T)}$ and $\ln(T/T_S)$ terms are negative, leading to $\Delta H_{TS-N(T)} > 0$. When $T = T_{S(\alpha)}$, $m_{TS-N(T)} = 0$, leading to a unique scenario wherein $\Delta G_{TS-N(T)} = \Delta H_{TS-N(T)} = \Delta S_{TS-N(T)} = 0$, and $k_{u(T)} = k^0$. Thus, $T_{S(\alpha)}$ is the first and the lowest temperature at which $\Delta H_{TS-N(T)} = 0$. In addition, the first extremum of $\Delta G_{TS-N(T)}$ ($\partial \Delta G_{TS-N(T)}/\partial T = -\Delta S_{TS-N(T)} = 0$), the first extremum of $\Delta G_{TS-N(T)}/T$ (i.e., the first derivative of the Massieu-Planck activation potential for unfolding is zero, or $\partial\left(\Delta G_{\text{TS-N}(T)}/T\right)/\partial T=-\Delta H_{\text{TS-N}(T)}/T^2=0$), and the first extremum of $k_{u(T)}$ ($\partial\ln k_{u(T)}/\partial T=\Delta H_{\text{TS-N}(T)}/RT^2=0$) occur at $T_{S(\alpha)}$; and while $\Delta G_{\text{TS-N}(T)}$ and $\Delta G_{\text{TS-N}(T)}/T$ are both a minimum, $k_{u(T)}$ is a maximum. Because $m_{\text{TS-N}(T)} > 0$ for $T_{S(\alpha)} < T < T_{S(\omega)}$, and $\ln(T/T_S) < 0$ for $T < T_S$, as long as $T\Delta C_{p\text{D-N}}\ln(T/T_S)/m_{\text{TS-N}(T)}\sqrt{\varphi} < -1$, $\Delta H_{\text{TS-N}(T)}$ will be negative. Since for $T < T_S$, $\ln(T/T_S)$ becomes less negative with a rise in temperature, at some point $T\Delta C_{p\text{D-N}}\ln(T/T_S)/m_{\text{TS-N}(T)}\sqrt{\varphi} = -1$ (or when $T\ln(T_S/T) = m_{\text{TS-N}(T)}\sqrt{\varphi}/\Delta C_{p\text{D-N}}$). Naturally, at this temperature the algebraic sum of the terms inside the square bracket on the RHS of Eq. (18) is zero leading to $\Delta H_{\text{TS-N}(T)}$ becoming zero for the second time. If we designate the temperature at which this occurs as $T_{H(\text{TS-N})}$, then $k_{u(T)}$ is given by

$$k_{u(T)}\Big|_{T=T_{H(\text{TS-N})}} = k^0 \exp\left(-\frac{\Delta G_{\text{TS-N}(T)}}{RT}\right)\Bigg|_{T=T_{H(\text{TS-N})}} = k^0 \exp\left(\frac{\Delta S_{\text{TS-N}(T)}}{R}\right)\Bigg|_{T=T_{H(\text{TS-N})}} \quad (19)$$

Substituting Eq. (11) in (19) gives

$$k_{u(T)}\Big|_{T=T_{H(\text{TS-N})}} = k^0 \exp\left(\frac{\omega\, m_{\text{TS-N}(T)}\Delta C_{p\text{D-N}}}{R\sqrt{\varphi}}\ln\left(\frac{T}{T_S}\right)\right)\Bigg|_{T=T_{H(\text{TS-N})}} \quad (20)$$

Because $\partial\ln k_{u(T)}/\partial T=\Delta H_{\text{TS-N}(T)}/RT^2=0$, a corollary is that the second extremum of $k_{u(T)}$ must occur when $T = T_{H(\text{TS-N})}$. In addition, since the first derivative of the Massieu-Planck activation potential for unfolding must also be zero, i.e., $\partial\left(\Delta G_{\text{TS-N}(T)}/T\right)/\partial T = -\Delta H_{\text{TS-N}(T)}/T^2 = 0$ when $T = T_{H(\text{TS-N})}$, and since $\Delta G_{\text{TS-N}(T)} = \omega\left(m_{\text{TS-N}(T)}\right)^2 > 0$ for $T \neq T_{S(\alpha)}$ and $T \neq T_{S(\omega)}$, and is a maximum when $T = T_s$, the conclusion is that $\Delta G_{\text{TS-N}(T)}/T$ must be a maximum at $T_{H(\text{TS-N})}$. Because $k_{u(T)}$ must be a minimum when $\Delta G_{\text{TS-N}(T)}/T$ is a maximum, a corollary is that for two-state systems at constant pressure and solvent conditions, "$k_{u(T)}$ is a minimum when the Gibbs barrier to unfolding is purely entropic." Importantly, in contrast to the first extrema of $k_{u(T)}$, $\Delta G_{\text{TS-N}(T)}$ and $\Delta G_{\text{TS-N}(T)}/T$ which occur at $T_{S(\alpha)}$, and are characterised by $k_{u(T)}$ being a maximum ($k_{u(T)} = k^0$), and both $\Delta G_{\text{TS-N}(T)}$ and $\Delta G_{\text{TS-N}(T)}/T$ being a minimum, the second extrema of $k_{u(T)}$ and $\Delta G_{\text{TS-N}(T)}/T$ which occur at $T_{H(\text{TS-N})}$ are characterised by $k_{u(T)}$ being a minimum and $\Delta G_{\text{TS-N}(T)}/T$ being a maximum (the reader will note that the second extremum of $\Delta G_{\text{TS-N}(T)}$ which is a maximum occurs not at $T_{H(\text{TS-N})}$ but at $T_S$). Now, the obvious question is: Where is $T_{H(\text{TS-N})}$ relative to $T_{S(\alpha)}$ and $T_S$? Because $\Delta G_{\text{TS-N}(T)} = \omega(m_{\text{TS-N}(T)})^2 > 0$ for $T \neq T_{S(\alpha)}$ and $T \neq T_{S(\omega)}$, $\Delta G_{\text{TS-N}(T)} = -T\Delta S_{\text{TS-N}(T)} > 0$ when $T = T_{H(\text{TS-N})}$; and the only way this condition can be satisfied is if $\Delta S_{\text{TS-N}(T)} < 0$. Since $\Delta S_{\text{TS-N}(T)} < 0$ only for $T_{S(\alpha)} < T < T_S$ (Eq. (11)), the logical conclusion is that $T_{H(\text{TS-N})}$ must lie between $T_{S(\alpha)}$ and $T_S$, or $T_{S(\alpha)} < T_{H(\text{TS-N})} < T_S$. When $T = T_S$ Eq. (18) reduces to $\Delta H_{\text{TS-N}(T)} = \omega(m_{\text{TS-N}(T)})^2 = \Delta G_{\text{TS-N}(T)} > 0$, i.e., the Gibbs barrier to unfolding is purely enthalpic at $T_S$, with $k_{u(T)}$ being given by

$$k_{u(T)}\big|_{T=T_S} = k^0 \exp\left(-\frac{\Delta G_{\text{TS-N}(T)}}{RT}\right)\bigg|_{T=T_S} = k^0 \exp\left(-\frac{\Delta H_{\text{TS-N}(T)}}{RT}\right)\bigg|_{T=T_S} \qquad (21)$$

Now, since $m_{\text{TS-N}(T)} > 0$ for $T_{S(\alpha)} < T < T_{S(\omega)}$, and $\ln(T/T_S) > 0$ for $T > T_S$, $\Delta H_{\text{TS-N}(T)} > 0$ for $T > T_S$. However, since $m_{\text{TS-N}(T)}$ is a maximum at $T_S$ and decreases with any deviation in temperature, the $\Delta H_{\text{TS-N}(T)}$ function initially increases with an increase in temperature, reaches saturation, and decreases with further rise in temperature; and precisely when $T = T_{S(\omega)}$, $m_{\text{TS-N}(T)} = 0$ and $\Delta H_{\text{TS-N}(T)} = 0$ for the third and the final time. Consequently, we may state that $\Delta H_{\text{TS-N}(T)} > 0$ for $T_{H(\text{TS-N})} < T < T_{S(\omega)}$. Akin to $T_{S(\alpha)}$, when $T = T_{S(\omega)}$, we note that $\Delta G_{\text{TS-N}(T)} = \Delta H_{\text{TS-N}(T)} = \Delta S_{\text{TS-N}(T)} = 0$, and $k_{u(T)} = k^0$. This is also the temperature at which both $\Delta G_{\text{TS-N}(T)}$ and $\Delta G_{\text{TS-N}(T)}/T$ are a minimum, and $k_{u(T)}$ is a maximum (the third and the final extremum). With further rise in temperature, i.e., for $T > T_{S(\omega)}$, $m_{\text{TS-N}(T)} < 0$ and $\ln(T/T_S) > 0$, leading to $\Delta H_{\text{TS-N}(T)} < 0$.

To summarize, we have three interesting scenarios: (*i*) $\Delta H_{\text{TS-N}(T)} > 0$ for $T_\alpha \leq T < T_{S(\alpha)}$ and $T_{H(\text{TS-N})} < T < T_{S(\omega)}$; (*ii*) $\Delta H_{\text{TS-N}(T)} < 0$ for $T_{S(\alpha)} < T < T_{H(\text{TS-N})}$ and $T_{S(\omega)} < T \leq T_\omega$; and (*iii*) $\Delta H_{\text{TS-N}(T)} = 0$ at $T_{S(\alpha)}$, $T_{H(\text{TS-N})}$, and $T_{S(\omega)}$. Thus, the activation of the native conformers to the TSE is enthalpically: (*i*) unfavourable for $T_\alpha \leq T < T_{S(\alpha)}$ and $T_{H(\text{TS-N})} < T < T_{S(\omega)}$; (*ii*) favourable for $T_{S(\alpha)} < T < T_{H(\text{TS-N})}$ and $T_{S(\omega)} < T \leq T_\omega$; and (*iii*) neither favourable nor unfavourable at $T_{S(\alpha)}$, $T_{H(\text{TS-N})}$, and $T_{S(\omega)}$. If we now reverse the direction of the reaction (i.e., for the partial folding

Page **15** of 45

reaction $TS \rightleftharpoons N$), we may state that the flux of the conformers from the TSE to the NSE is enthalpically: (*i*) favourable for $T_\alpha \leq T < T_{S(\alpha)}$ and $T_{H(\text{TS-N})} < T < T_{S(\omega)}$ ($\Delta H_{\text{TS-N}(T)} > 0 \Rightarrow \Delta H_{\text{N-TS}(T)} < 0$); (*ii*) unfavourable for $T_{S(\alpha)} < T < T_{H(\text{TS-N})}$ and $T_{S(\omega)} < T \leq T_\omega$ ($\Delta H_{\text{TS-N}(T)} < 0 \Rightarrow \Delta H_{\text{N-TS}(T)} > 0$); and (*iii*) neither favourable nor unfavourable at $T_{S(\alpha)}$, $T_{H(\text{TS-N})}$, and $T_{S(\omega)}$. In addition, $k_{u(T)} < k^0$ for $T \neq T_{S(\alpha)}$ and $T \neq T_{S(\omega)}$, $k_{u(T)} = k^0$ when $T = T_{S(\alpha)}$ and $T = T_{S(\omega)}$, and a minimum when $T = T_{H(\text{TS-N})}$. Further, while protein *unfolding* occurs *via* a conventional barrier-limited process for the temperature regime $T_{S(\alpha)} < T < T_{S(\omega)}$, and is barrierless when $T = T_{S(\alpha)}$ and $T = T_{S(\omega)}$, the ultralow temperature regime $T_\alpha \leq T < T_{S(\alpha)}$, and the high temperature regime $T_{S(\omega)} < T \leq T_\omega$ are once again barrier-limited but fall under the *Marcus-inverted-regime*.[25]

**Barrierless and Marcus-inverted-regimes**

From the perspective of the parabolic hypothesis, *barrierless folding* is characterised by the condition: $m_{\text{TS-D}(T)} = 0 \Rightarrow \alpha \left( m_{\text{TS-D}(T)} \right)^2 = \Delta G_{\text{TS-D}(T)} = 0$. Consequently, for a barrierless folder $\Delta G_{\text{D-N}(T)} = \Delta G_{\text{TS-N}(T)}$ and $m_{\text{TS-N}(T)} = m_{\text{D-N}}$. In a parabolic representation, the left-arm of the NSE-parabola intersects the vertex of the DSE-parabola. For the special case of the *Marcus-inverted-regime* associated with the folding reaction, the *curve-crossing* occurs to the left of the vertex of the DSE-parabola, i.e., the left-arm of the NSE-parabola intersects the left-arm of the DSE-parabola leading to $m_{\text{TS-N}(T)} > m_{\text{D-N}}$ (see Figure 6 in Marcus, 1993).[25] In contrast, *barrierless unfolding* is characterised by the condition: $m_{\text{TS-N}(T)} = 0 \Rightarrow \omega \left( m_{\text{TS-N}(T)} \right)^2 = \Delta G_{\text{TS-N}(T)} = 0$, leading to $\Delta G_{\text{D-N}(T)} = -\Delta G_{\text{TS-D}(T)} = -\alpha \left( m_{\text{TS-D}(T)} \right)^2$; and because $m_{\text{TS-D}(T)} = m_{\text{D-N}}$ for this scenario, it implies $\Delta G_{\text{D-N}(T)} = -\alpha \left( m_{\text{D-N}} \right)^2 = -\lambda$. In a parabolic representation, the right-arm of the DSE-parabola intersects the vertex of the NSE-parabola. For the special case of the *Marcus-inverted-regime* associated with the unfolding reaction, the *curve-crossing* occurs to the right of the vertex of the NSE-parabola (i.e., the right-arm of the DSE-parabola intersects the right-arm of the NSE-parabola) leading to $m_{\text{TS-D}(T)} > m_{\text{D-N}}$.

Thus, although from the perspective of the parabolic hypothesis *barrierless unfolding* and the *Marcus-inverted-regimes* associated with the unfolding reaction are a theoretical possibility for two-state-folding proteins, as long as $m_{TS-D(T)} \neq 0$ and $m_{TS-N(T)} \neq 0$, the protein cannot fold or unfold, respectively, *via* downhill mechanism, irrespective of whether or not it is an ultrafast folder. Because the extremum of $m_{TS-D(T)}$ which is a minimum occurs at $T_S$, it essentially implies that as long as there exists a chevron with a well-defined linear folding arm at $T_S$, it is theoretically impossible for a two-state folder at constant pressure and solvent conditions to spontaneously (i.e., unaided by ligands, co-solvents etc.) switch to a downhill mechanism, no matter what the temperature. A corollary is that "*chevrons with well-defined linear folding and unfolding arms are fundamentally incompatible with downhill scenarios.*"[26] The reader will note that the theoretically impossible downhill folding scenario that is being referred to here is not the one wherein the denatured conformers spontaneously fold to their native states *via* a *first-order* process with $k_{f(T)} \cong k^0$ (manifest when $\Delta G_{TS-D(T)}$ is approximately equal to thermal noise, i.e., $\Delta G_{TS-D(T)} < \cong 3RT$), but the controversial Type 0 scenario according to the Energy Landscape Theory, (see Figure 6 in Onuchic et al., 1997) wherein the conformers in the DSE ostensibly reach the NSE without encountering any barrier ($\Delta G_{TS-D(T)} = 0$).[27-30] A detailed discussion on barrierless folding is beyond the scope of this article and will be addressed in subsequent publications.

**Determinants of Gibbs activation energies for folding and unfolding**

The determinants of $\Delta G_{TS-D(T)}$ in terms of its activation enthalpy and entropy may be readily deduced by partitioning the entire temperature range over which the two-state system is physically defined ($T_\alpha \leq T \leq T_\omega$) into three distinct regimes using four unique reference temperatures: $T_\alpha$, $T_S$, $T_{H(TS-D)}$, and $T_\omega$.

1. For $T_\alpha \leq T < T_S$, the activation of conformers from the DSE to the TSE is entropically favoured ($T\Delta S_{TS-D(T)} > 0$) but is more than offset by the endothermic activation enthalpy ($\Delta H_{TS-D(T)} > 0$), leading to incomplete compensation and a positive $\Delta G_{TS-D(T)}$ ($\Delta H_{TS-D(T)} - T\Delta S_{TS-D(T)} > 0$). When $T = T_S$, the positive Gibbs barrier to folding is purely due to the endothermic enthalpy of activation ($\Delta G_{TS-D(T)} = \Delta H_{TS-D(T)} > 0$) and $\Delta G_{TS-D(T)}$ is a minimum (its lone extremum).

2. For $T_S < T < T_{H(TS-D)}$, the activation of denatured conformers to the TSE is enthalpically and entropically disfavoured ($\Delta H_{TS-D(T)} > 0$ and $T\Delta S_{TS-D(T)} < 0$) leading to a positive $\Delta G_{TS-D(T)}$.

3. In contrast, for $T_{H(TS-D)} < T \leq T_\omega$, the favourable exothermic activation enthalpy for folding ($\Delta H_{TS-D(T)} < 0$) is more than offset by the unfavourable entropy of activation for folding ($T\Delta S_{TS-D(T)} < 0$), leading once again to a positive $\Delta G_{TS-D(T)}$. When $T = T_{H(TS-D)}$, $\Delta G_{TS-D(T)}$ is purely due to the negative change in the activation entropy or the *negentropy* of activation ( $\Delta G_{TS-D(T)} = -T\Delta S_{TS-D(T)} > 0$ ), $\Delta G_{TS-D(T)}/T$ is a minimum and $k_{f(T)}$ is a maximum (their lone extrema).

An important conclusion that we may draw from these analyses is the following: While it is true that for the temperature regimes $T_\alpha \leq T < T_S$ and $T_{H(TS-D)} < T \leq T_\omega$, the positive Gibbs barrier to folding is due to the incomplete compensation of the opposing activation enthalpy and entropy, this is clearly not the case for $T_S < T < T_{H(TS-D)}$ where both these two state functions are unfavourable and collude to yield a positive Gibbs activation barrier. In short, the Gibbs barrier to folding is not always due to the incomplete compensation of the opposing enthalpy and entropy.

Similarly, the determinants of $\Delta G_{TS-N(T)}$ in terms of its activation enthalpy and entropy may be readily divined by partitioning the entire temperature range into five distinct regimes using six unique reference temperatures: $T_\alpha$, $T_{S(\alpha)}$, $T_{H(TS-N)}$, $T_S$, $T_{S(\omega)}$, and $T_\omega$.

1. For $T_\alpha \leq T < T_{S(\alpha)}$, which is the ultralow temperature *Marcus-inverted-regime* for protein unfolding, the activation of the native conformers to the TSE is entropically favoured ($T\Delta S_{TS-N(T)} > 0$) but is more than offset by the unfavourable enthalpy of activation ($\Delta H_{TS-N(T)} > 0$) leading to incomplete compensation and a positive $\Delta G_{TS-N(T)}$ ($\Delta H_{TS-N(T)} - T\Delta S_{TS-N(T)} > 0$). When $T = T_{S(\alpha)}$, $\Delta S_{TS-N(T)} = \Delta H_{TS-N(T)} = 0 \Rightarrow \Delta G_{TS-N(T)} = 0$. The first extrema of $\Delta G_{TS-N(T)}$ and $\Delta G_{TS-N(T)}/T$ (which are a minimum), and the first extremum of $k_{u(T)}$ (which is a maximum, $k_{u(T)} = k^0$) occur at $T_{S(\alpha)}$.

2. For $T_{S(\alpha)} < T < T_{H(TS-N)}$, the activation of the native conformers to the TSE is enthalpically favourable ($\Delta H_{TS-N(T)} < 0$) but is more than offset by the unfavourable negentropy of activation ($T\Delta S_{TS-N(T)} < 0$) leading to $\Delta G_{TS-N(T)} > 0$. When $T = T_{H(TS-N)}$, $\Delta H_{TS-N(T)} = 0$ for the second time, and the Gibbs barrier to unfolding is purely due to the negentropy of activation ($\Delta G_{\text{TS-N}(T)} = -T\Delta S_{\text{TS-N}(T)} > 0$). The second extrema of $\Delta G_{\text{TS-N}(T)}/T$ (which is a maximum) and $k_{u(T)}$ (which is a minimum) occur at $T_{H(\text{TS-N})}$.

3. For $T_{H(\text{TS-N})} < T < T_S$, the activation of the native conformers to the TSE is entropically and enthalpically unfavourable ($\Delta H_{\text{TS-N}(T)} > 0$ and $T\Delta S_{\text{TS-N}(T)} < 0$) leading to $\Delta G_{\text{TS-N}(T)} > 0$. When $T = T_S$, $\Delta S_{\text{TS-N}(T)} = 0$ for the second time, and the Gibbs barrier to unfolding is purely due to the endothermic enthalpy of activation ($\Delta G_{\text{TS-N}(T)} = \Delta H_{\text{TS-N}(T)} > 0$). The second extremum of $\Delta G_{\text{TS-N}(T)}$ (which is a maximum) occurs at $T_S$.

4. For $T_S < T < T_{S(\omega)}$, the activation of the native conformers to the TSE is entropically favourable ($T\Delta S_{\text{TS-N}(T)} > 0$) but is more than offset by the endothermic enthalpic of activation ($\Delta H_{\text{TS-N}(T)} > 0$) leading to incomplete compensation and a positive $\Delta G_{\text{TS-N}(T)}$. When $T = T_{S(\omega)}$, $\Delta S_{\text{TS-N}(T)} = \Delta H_{\text{TS-N}(T)} = 0$ for the third and the final time, and $\Delta G_{\text{TS-N}(T)} = 0$ for the second and final time. The third extrema of $\Delta G_{\text{TS-N}(T)}$ and $\Delta G_{\text{TS-N}(T)}/T$ (which are a minimum), and the third extremum of $k_{u(T)}$ (which is a maximum, $k_{u(T)} = k^0$) occur at $T_{S(\omega)}$.

5. For $T_{S(\omega)} < T \leq T_\omega$, which is the high temperature *Marcus-inverted-regime* for protein unfolding, the activation of the native conformers to the TSE is enthalpically favourable ($\Delta H_{\text{TS-N}(T)} < 0$) but is more than offset by the unfavourable negentropy of activation ($T\Delta S_{\text{TS-N}(T)} < 0$), leading to $\Delta G_{\text{TS-N}(T)} > 0$.

Once again we note that although the Gibbs barrier to unfolding is due to the incomplete compensation of the opposing enthalpies and entropies of activation for the temperature regimes $T_\alpha \leq T < T_{S(\alpha)}$, $T_{S(\alpha)} < T < T_{H(\text{TS-N})}$, $T_S < T < T_{S(\omega)}$, and $T_{S(\omega)} < T \leq T_\omega$, both the enthalpy and the entropy of activation are unfavourable and collude to generate the Gibbs barrier to unfolding for the temperature regime $T_{H(\text{TS-N})} < T < T_S$.

However, in a *protein folding* scenario where the activated conformers diffuse on the Gibbs energy surface to reach the NSE, the interpretation changes because the algebraic signs of the state functions invert. Thus, for the reaction $[TS] \rightleftharpoons N$ we may state:

1. For $T_\alpha \leq T < T_{S(\alpha)}$, the flux of the conformers from the TSE to the NSE is entropically disfavoured ($T\Delta S_{\text{TS-N}(T)} > 0 \Rightarrow T\Delta S_{\text{N-TS}(T)} < 0$) but is more than compensated by the favourable change in enthalpy ($\Delta H_{\text{TS-N}(T)} > 0 \Rightarrow \Delta H_{\text{N-TS}(T)} < 0$), leading to $\Delta G_{\text{N-TS}(T)} < 0$.

2. For $T_{S(\alpha)} < T < T_{H(\text{TS-N})}$, the flux of the conformers from the TSE to the NSE is enthalpically unfavourable ($\Delta H_{\text{TS-N}(T)} < 0 \Rightarrow \Delta H_{\text{N-TS}(T)} > 0$) but is more than compensated by the favourable change in entropy ($T\Delta S_{\text{TS-N}(T)} < 0 \Rightarrow T\Delta S_{\text{N-TS}(T)} > 0$) leading to $\Delta G_{\text{N-TS}(T)} < 0$. When $T = T_{H(\text{TS-N})}$, the flux is driven purely by the positive change in entropy ($\Delta G_{\text{N-TS}(T)} = -T\Delta S_{\text{N-TS}(T)} < 0$).

3. For $T_{H(\text{TS-N})} < T < T_S$, the flux of the conformers from the TSE to the NSE is entropically and enthalpically favourable ($\Delta H_{\text{N-TS}(T)} < 0$ and $T\Delta S_{\text{N-TS}(T)} > 0$) leading to $\Delta G_{\text{N-TS}(T)} < 0$. When $T = T_S$, the flux is driven purely by the exothermic change in enthalpy ($\Delta G_{\text{N-TS}(T)} = \Delta H_{\text{N-TS}(T)} < 0$).

4. For $T_S < T < T_{S(\omega)}$, the flux of the conformers from the TSE to the NSE is entropically unfavourable ($T\Delta S_{\text{TS-N}(T)} > 0 \Rightarrow T\Delta S_{\text{N-TS}(T)} < 0$) but is more than compensated by the exothermic change in enthalpy ($\Delta H_{\text{TS-N}(T)} > 0 \Rightarrow \Delta H_{\text{N-TS}(T)} < 0$) leading to $\Delta G_{\text{N-TS}(T)} < 0$.

5. For $T_{S(\omega)} < T \leq T_\omega$, the flux of the conformers from the TSE to the NSE is enthalpically unfavourable ($\Delta H_{\text{TS-N}(T)} < 0 \Rightarrow \Delta H_{\text{N-TS}(T)} > 0$) but is more than compensated by the favourable change in entropy ($T\Delta S_{\text{TS-N}(T)} < 0 \Rightarrow T\Delta S_{\text{N-TS}(T)} > 0$), leading to $\Delta G_{\text{N-TS}(T)} < 0$.

A detailed analysis of the determinants of the Gibbs activation energies in terms of the *chain and desolvation enthalpies*, and the *chain and desolvation entropies* is beyond the scope of this article and will be addressed in subsequent publications.

## Heat capacities of activation for folding and unfolding

The heat capacity of activation for partial folding reaction $D \rightleftharpoons [TS]$ ($\Delta C_{p\text{TS-D}(T)}$) is given by the derivative of $\Delta S_{\text{TS-D}(T)}$ (Eq. (8)) with respect to temperature.

$$\frac{\partial \Delta S_{\text{TS-D}(T)}}{\partial T} = \frac{\Delta C_{p\text{TS-D}(T)}}{T} = -\alpha \frac{\partial^2 \left(m_{\text{TS-D}(T)}\right)^2}{\partial T^2} \tag{22}$$

The solution for Eq. (22) is given by (see **Appendix**)

$$\Delta C_{p\text{TS-D}(T)} = -\frac{\alpha}{2\varphi\sqrt{\varphi}}\left[m_{\text{TS-D}(T)}2\varphi\Delta C_{p\text{D-N}} + \omega m_{\text{D-N}}T\left(\Delta S_{\text{D-N}(T)}\right)^2\right] \quad (23)$$

Because a negative change in heat capacity is less intuitive, we may recast Eq. (23) by changing the reaction-direction (i.e., $[TS] \rightleftharpoons D$) to give

$$\Delta C_{p\text{D-TS}(T)} = \frac{\alpha}{2\varphi\sqrt{\varphi}}\left[m_{\text{TS-D}(T)}2\varphi\Delta C_{p\text{D-N}} + \omega m_{\text{D-N}}T\left(\Delta S_{\text{D-N}(T)}\right)^2\right] \quad (24)$$

Substituting Tanford's relationship $m_{\text{TS-D}(T)} = \beta_{T(\text{fold})(T)}m_{\text{D-N}}$ in Eq. (24) and recasting gives[31]

$$\Delta C_{p\text{D-TS}(T)} = \frac{\alpha m_{\text{D-N}}}{2\varphi\sqrt{\varphi}}\left[\beta_{T(\text{fold})(T)}2\varphi\Delta C_{p\text{D-N}} + \omega T\left(\Delta S_{\text{D-N}(T)}\right)^2\right] \quad (25)$$

The heat capacity of activation for the partial unfolding reaction $N \rightleftharpoons [TS]$ ($\Delta C_{p\text{TS-N}(T)}$) is given by the derivative of $\Delta S_{\text{TS-N}(T)}$ (Eq. (10)) with respect to temperature.

$$\frac{\partial \Delta S_{\text{TS-N}(T)}}{\partial T} = \frac{\Delta C_{p\text{TS-N}(T)}}{T} = -\omega\frac{\partial^2\left(m_{\text{TS-N}(T)}\right)^2}{\partial T^2} \quad (26)$$

The solution for Eq. (26) is given by (see **Appendix**)

$$\Delta C_{p\text{TS-N}(T)} = \frac{\omega}{2\varphi\sqrt{\varphi}}\left[m_{\text{TS-N}(T)}2\varphi\Delta C_{p\text{D-N}} - \alpha m_{\text{D-N}}T\left(\Delta S_{\text{D-N}(T)}\right)^2\right] \quad (27)$$

Substituting $m_{\text{TS-N}(T)} = \beta_{T(\text{unfold})(T)}m_{\text{D-N}}$ in Eq. (27) and recasting gives

$$\Delta C_{p\text{TS-N}(T)} = \frac{\omega m_{\text{D-N}}}{2\varphi\sqrt{\varphi}}\left[\beta_{T(\text{unfold})(T)}2\varphi\Delta C_{p\text{D-N}} - \alpha T\left(\Delta S_{\text{D-N}(T)}\right)^2\right] \quad (28)$$

Although not shown, identical expressions for the heat capacities of activation for folding and unfolding may be obtained by differentiating the activation enthalpies for folding and unfolding with respect to temperature.

It is instructive to further analyse Eqs. (24) and (27): If we recall that the force constants and $m_{\text{D-N}}$ are temperature-invariant, it becomes readily apparent that the second terms in the square brackets on the right-hand-side (RHS) i.e., $\omega T\left(\Delta S_{\text{D-N}(T)}\right)^2$ and $\alpha T\left(\Delta S_{\text{D-N}(T)}\right)^2$ will be parabolas, and their values positive for $T \neq T_S$, and zero when $T = T_S$. This is due to $\Delta S_{\text{D-N}(T)}$ being negative for $T < T_S$, positive for $T > T_S$, and zero for $T = T_S$.[4] Furthermore, we note that $\varphi$, $\sqrt{\varphi}$ and $m_{\text{TS-N}(T)}$ are a maximum, and $m_{\text{TS-D}(T)}$ a minimum at $T_S$. Consequently, for a two-state folder at constant pressure and solvent conditions, $\Delta C_{p\text{D-TS}(T)}$ is a minimum (or $\Delta C_{p\text{TS-D}(T)}$ is the least negative), and $\Delta C_{p\text{TS-N}(T)}$ is a maximum when $T = T_S$. Thus, Eqs. (24) and (27) become

$$\Delta C_{p\text{D-TS}(T)}\Big|_{T=T_S} = \frac{\alpha m_{\text{TS-D}(T)} \Delta C_{p\text{D-N}}}{\sqrt{\varphi}}\Bigg|_{T=T_S} > 0 \qquad (29)$$

$$\Delta C_{p\text{TS-N}(T)}\Big|_{T=T_S} = \frac{\omega m_{\text{TS-N}(T)} \Delta C_{p\text{D-N}}}{\sqrt{\varphi}}\Bigg|_{T=T_S} > 0 \qquad (30)$$

Since $\Delta S_{\text{TS-D}(T)}$ and $\Delta S_{\text{TS-N}(T)}$ are zero, $\Delta G_{\text{D-N}(T)}$ and $\Delta G_{\text{TS-N}(T)}$ are a maximum, and $\Delta G_{\text{TS-D}(T)}$ is a minimum when $T = T_S$, a corollary is that the *Gibbs barriers to folding and unfolding are a minimum and a maximum, respectively, and equilibrium stability is a maximum, and are all purely enthalpic when $\Delta C_{p\text{D-TS}(T)}$ and $\Delta C_{p\text{TS-N}(T)}$ are a minimum and a maximum, respectively.* Consistent with the molecular interpretation of change in heat capacity, $\Delta C_{p\text{D-TS}(T)}$ is a minimum when the conformers in the DSE travel the least distance from the mean SASA of their ensemble along the SASA-RC to reach the TSE (see the *principle of least displacement* above), and $\Delta C_{p\text{TS-N}(T)}$ is a maximum when the conformers in the NSE expose the greatest amount of SASA to reach the TSE. See **Appendix** for what become of Eqs. (24) and (27) at $T_{S(\alpha)}$ and $T_{S(\omega)}$, and their implications.

## Comparison of RCs

### Heat capacity RC

Adopting Leffler's framework, the relative sensitivities of the activation and equilibrium enthalpies in response to a perturbation in temperature is given by the ratio of the derivatives of the activation and equilibrium enthalpies with respect to temperature.[5,32] Thus, for the partial folding reaction $D \rightleftharpoons [TS]$, we have

$$\beta_{H(\text{fold})(T)} = \frac{\partial \Delta H_{\text{TS-D}(T)}/\partial T}{\partial \Delta H_{\text{N-D}(T)}/\partial T} = \frac{\Delta C_{p\text{TS-D}(T)}}{\Delta C_{p\text{N-D}}} = \frac{\Delta C_{p\text{D-TS}(T)}}{\Delta C_{p\text{D-N}}} \tag{31}$$

where $\beta_{H(\text{fold})(T)}$ is classically interpreted to be a measure of the position of the TSE relative to the DSE along the heat capacity RC. Similarly, for the partial unfolding reaction $N \rightleftharpoons [TS]$ we may write

$$\beta_{H(\text{unfold})(T)} = \frac{\partial \Delta H_{\text{TS-N}(T)}/\partial T}{\partial \Delta H_{\text{D-N}(T)}/\partial T} = \frac{\Delta C_{p\text{TS-N}(T)}}{\Delta C_{p\text{D-N}}} = \frac{\Delta C_{p\text{N-TS}(T)}}{\Delta C_{p\text{N-D}}} \tag{32}$$

where $\beta_{H(\text{unfold})(T)}$ measure of the position of the TSE relative to the NSE along the heat capacity RC. Naturally, for a two-state system $\beta_{H(\text{fold})(T)} + \beta_{H(\text{unfold})(T)} = 1$ for any given reaction-direction.

Similarly, the relative sensitivities of the activation and equilibrium entropies to a perturbation in temperature are given by the ratio of the derivatives of the activation and equilibrium entropies with respect to temperature. Thus, we may write

$$\beta_{S(\text{fold})(T)} = \frac{\partial \Delta S_{\text{TS-D}(T)}/\partial T}{\partial \Delta S_{\text{N-D}(T)}/\partial T} = \frac{\Delta C_{p\text{TS-D}(T)}/T}{\Delta C_{p\text{N-D}}/T} = \frac{\Delta C_{p\text{TS-D}(T)}}{\Delta C_{p\text{N-D}}} = \frac{\Delta C_{p\text{D-TS}(T)}}{\Delta C_{p\text{D-N}}} \tag{33}$$

$$\beta_{S(\text{unfold})(T)} = \frac{\partial \Delta S_{\text{TS-N}(T)}/\partial T}{\partial \Delta S_{\text{D-N}(T)}/\partial T} = \frac{\Delta C_{p\text{TS-N}(T)}/T}{\Delta C_{p\text{D-N}}/T} = \frac{\Delta C_{p\text{TS-N}(T)}}{\Delta C_{p\text{D-N}}} = \frac{\Delta C_{p\text{N-TS}(T)}}{\Delta C_{p\text{N-D}}} \tag{34}$$

where $\beta_{S(\text{fold})(T)}$ and $\beta_{S(\text{unfold})(T)}$ are identical to $\beta_{H(\text{fold})(T)}$ and $\beta_{H(\text{unfold})(T)}$, respectively (compare Eqs. (33) and (34) with (31) and (32), respectively). Dividing Eqs. (24) and (27) by $\Delta C_{p\text{D-N}}$ gives

$$\begin{aligned}\beta_{H(\text{fold})(T)} &= \frac{\alpha}{2\varphi\sqrt{\varphi}\Delta C_{p\text{D-N}}}\left[m_{\text{TS-D}(T)}2\varphi\Delta C_{p\text{D-N}} + \omega m_{\text{D-N}}T\left(\Delta S_{\text{D-N}(T)}\right)^2\right] \\ &= \frac{\alpha m_{\text{D-N}}}{2\varphi\sqrt{\varphi}\Delta C_{p\text{D-N}}}\left[\beta_{T(\text{fold})(T)}2\varphi\Delta C_{p\text{D-N}} + \omega T\left(\Delta S_{\text{D-N}(T)}\right)^2\right]\end{aligned} \tag{35}$$

$$\begin{aligned}\beta_{H(\text{unfold})(T)} &= \frac{\omega}{2\varphi\sqrt{\varphi}\Delta C_{p\text{D-N}}}\left[m_{\text{TS-N}(T)}2\varphi\Delta C_{p\text{D-N}} - \alpha m_{\text{D-N}}T\left(\Delta S_{\text{D-N}(T)}\right)^2\right] \\ &= \frac{\omega m_{\text{D-N}}}{2\varphi\sqrt{\varphi}\Delta C_{p\text{D-N}}}\left[\beta_{T(\text{unfold})(T)}2\varphi\Delta C_{p\text{D-N}} - \alpha T\left(\Delta S_{\text{D-N}(T)}\right)^2\right]\end{aligned} \tag{36}$$

When $T = T_S$, $\Delta S_{\text{D-N}(T)} = 0$ and Eqs. (35) and (36) reduce to

$$\beta_{\text{H(fold)}(T)}\Big|_{T=T_S} \equiv \beta_{\text{S(fold)}(T)}\Big|_{T=T_S} = \frac{\alpha m_{\text{TS-D}(T)}}{\sqrt{\varphi}}\Bigg|_{T=T_S} = \frac{\alpha \beta_{\text{T(fold)}(T)} m_{\text{D-N}}}{\sqrt{\varphi}}\Bigg|_{T=T_S} \tag{37}$$

$$\beta_{\text{H(unfold)}(T)}\Big|_{T=T_S} \equiv \beta_{\text{S(unfold)}(T)}\Big|_{T=T_S} = \frac{\omega m_{\text{TS-N}(T)}}{\sqrt{\varphi}}\Bigg|_{T=T_S} = \frac{\omega \beta_{\text{T(unfold)}(T)} m_{\text{D-N}}}{\sqrt{\varphi}}\Bigg|_{T=T_S} \tag{38}$$

As explained earlier, because $\Delta C_{p\text{D-N}}$ is temperature-invariant by postulate, and $\Delta C_{p\text{D-TS}(T)}$ is a minimum, and $\Delta C_{p\text{TS-N}(T)}$ is a maximum at $T_S$, $\beta_{\text{H(fold)}(T)}$ and $\beta_{\text{H(unfold)}(T)}$ are a minimum and a maximum, respectively, at $T_S$. How do $\beta_{\text{H(fold)}(T)}$ and $\beta_{\text{H(unfold)}(T)}$ compare with their counterparts, $\beta_{\text{T(fold)}(T)}$ and $\beta_{\text{T(unfold)}(T)}$? This is important because a statistically significant correlation exists between $m_{\text{D-N}}$ and $\Delta C_{p\text{D-N}}$, and both these two parameters independently correlate with $\Delta \text{SASA}_{\text{D-N}}$.[33,34] Recasting Eqs. (37) and (38) gives

$$\frac{\beta_{\text{H(fold)}(T)}}{\beta_{\text{T(fold)}(T)}}\Bigg|_{T=T_S} \equiv \frac{\beta_{\text{S(fold)}(T)}}{\beta_{\text{T(fold)}(T)}}\Bigg|_{T=T_S} = \frac{\alpha m_{\text{D-N}}}{\sqrt{\varphi}}\Bigg|_{T=T_S} < 1 \tag{39}$$

$$\frac{\beta_{\text{H(unfold)}(T)}}{\beta_{\text{T(unfold)}(T)}}\Bigg|_{T=T_S} \equiv \frac{\beta_{\text{S(unfold)}(T)}}{\beta_{\text{T(unfold)}(T)}}\Bigg|_{T=T_S} = \frac{\omega m_{\text{D-N}}}{\sqrt{\varphi}}\Bigg|_{T=T_S} > 1 \tag{40}$$

Since $m_{\text{TS-N}(T)} > 0$ and a maximum, and $m_{\text{TS-D}(T)} > 0$ and a minimum, respectively at $T_S$, it is readily apparent from inspection of Eqs. (1) and (2) that $\sqrt{\varphi} > \alpha m_{\text{D-N}}$ and $\omega m_{\text{D-N}} > \sqrt{\varphi}$ at $T_S$. Consequently, $\beta_{\text{T(fold)}(T)}\Big|_{T=T_S} > \beta_{\text{H(fold)}(T)}\Big|_{T=T_S}$ and $\beta_{\text{T(unfold)}(T)}\Big|_{T=T_S} < \beta_{\text{H(unfold)}(T)}\Big|_{T=T_S}$. We will demonstrate using experimental data in subsequent publications that while $\beta_{\text{H(fold)}(T)}$ and $\beta_{\text{H(unfold)}(T)}$ have qualitatively similar dependences on temperature as do $\beta_{\text{T(fold)}(T)}$ and $\beta_{\text{T(unfold)}(T)}$, respectively, and both $\beta_{\text{H(fold)}(T)}$ and $\beta_{\text{T(fold)}(T)}$ are a minimum, and both $\beta_{\text{H(unfold)}(T)}$ and $\beta_{\text{T(unfold)}(T)}$ are a maximum, respectively, at $T_S$, $\beta_{\text{H(fold)}(T)} \neq \beta_{\text{T(fold)}(T)}$ and $\beta_{\text{H(unfold)}(T)} \neq \beta_{\text{T(unfold)}(T)}$ except for two unique temperatures, one in the low temperature, and the other in the high temperature regime, and that $\beta_{\text{H(fold)}(T)} < \beta_{\text{T(fold)}(T)}$ and $\beta_{\text{H(unfold)}(T)} > \beta_{\text{T(unfold)}(T)}$ across a wide temperature regime (see page 178 in Bilsel and Matthews, 2000).[35] This has certain implications for the origin of the large and positive heat capacity for protein unfolding at equilibrium and is addressed in subsequent articles.

**Entropic RC**

The Leffler parameters for the relative sensitivities of the activation and equilibrium Gibbs energies in response to a perturbation in temperature are given by the ratios of the derivatives of the activation and equilibrium Gibbs energies with respect to temperature.[7-9] Thus, for the partial folding reaction $D \rightleftharpoons [TS]$, we have

$$\beta_{G(\text{fold})(T)} = \frac{\partial \Delta G_{\text{TS-D}(T)}/\partial T}{\partial \Delta G_{\text{N-D}(T)}/\partial T} = \frac{-\Delta S_{\text{TS-D}(T)}}{-\Delta S_{\text{N-D}(T)}} = \frac{-\Delta S_{\text{TS-D}(T)}}{\Delta S_{\text{D-N}(T)}} \tag{41}$$

where $\beta_{G(\text{fold})(T)}$ is classically interpreted to be a measure of the position of the TSE relative to the DSE along the entropic RC. Substituting Eqs. (9) and (A8) in (41) and rearranging gives

$$\beta_{G(\text{fold})(T)} = \frac{\alpha\, m_{\text{TS-D}(T)}\, \cancel{\Delta S_{\text{D-N}(T)}}}{\cancel{\Delta S_{\text{D-N}(T)}}\, \sqrt{\varphi}} = \frac{\alpha\, m_{\text{TS-D}(T)}}{\sqrt{\varphi}} \tag{42}$$

Similarly for the partial unfolding reaction $N \rightleftharpoons [TS]$ we have

$$\beta_{G(\text{unfold})(T)} = \frac{\partial \Delta G_{\text{TS-N}(T)}/\partial T}{\partial \Delta G_{\text{D-N}(T)}/\partial T} = \frac{\Delta S_{\text{TS-N}(T)}}{\Delta S_{\text{D-N}(T)}} \tag{43}$$

where $\beta_{G(\text{unfold})(T)}$ is a measure of the position of the TSE relative to the NSE along the entropic RC. Substituting Eqs. (11) and (A8) in (43) gives

$$\beta_{G(\text{unfold})(T)} = \frac{\omega\, m_{\text{TS-N}(T)}\, \cancel{\Delta S_{\text{D-N}(T)}}}{\cancel{\Delta S_{\text{D-N}(T)}}\, \sqrt{\varphi}} = \frac{\omega\, m_{\text{TS-N}(T)}}{\sqrt{\varphi}} \tag{44}$$

Inspection of Eqs. (41) and (43) shows that $\beta_{G(\text{fold})(T)} + \beta_{G(\text{unfold})(T)} = 1$ for any given reaction-direction. Now, since $\Delta S_{\text{D-N}(T)} = \Delta S_{\text{TS-D}(T)} = \Delta S_{\text{TS-N}(T)} = 0$ at $T_S$, $\beta_{G(\text{fold})(T)}$ and $\beta_{G(\text{unfold})(T)}$ calculated using Eqs. (41) and (43), respectively, will be undefined or indeterminate for $T = T_S$, with plots of the temperature-dependence of the same having vertical asymptotes at $T_S$. However, these are removable point discontinuities as is apparent from Eqs. (42) and (44); consequently, graphs generated using the latter set of equations will be devoid of the vertical asymptotes, albeit with a hole at $T_S$. If we ignore the hole at $T_S$ to enable a physical description and their comparison to other RCs, the extremum of $\beta_{G(fold)(T)}$ (which is positive and a minimum) and the extremum of $\beta_{G(unfold)(T)}$ (which is positive and a maximum) will occur at $T_S$. This is because $m_{TS-D(T)}$ is a minimum, and $m_{TS-N(T)}$ and $\varphi$ are a maximum, respectively, at $T_S$. Further, since $m_{TS-N(T)} = 0$ at $T_{S(\alpha)}$ and $T_{S(\omega)}$, $\beta_{G(fold)(T)} \equiv \beta_{T(fold)(T)} = 1$ and $\beta_{G(unfold)(T)} \equiv \beta_{T(unfold)(T)} = 0$ at the same; and for $T_\alpha \leq T < T_{S(\alpha)}$ and $T_{S(\omega)} < T \leq T_\omega$ (the ultralow and high temperature *Marcus-inverted-regimes*, respectively), $\beta_{G(fold)(T)}$ and $\beta_{T(fold)(T)}$ are greater than unity, and $\beta_{G(unfold)(T)}$ and $\beta_{T(unfold)(T)}$ are negative. Comparison of Eqs. (37) and (42), and Eqs. (38) and (44) demonstrate that when $T = T_S$, we have

$$\beta_{H(fold)(T_S)} \equiv \beta_{S(fold)(T_S)} \equiv \beta_{G(fold)(T)} \tag{45}$$

$$\beta_{H(unfold)(T_S)} \equiv \beta_{S(unfold)(T_S)} \equiv \beta_{G(unfold)(T)} \tag{46}$$

Thus, the position of the TSE along the heat capacity and entropic RCs are identical at $T_S$, and non-identical for all $T \neq T_S$. In physical organic chemistry, the terms $\beta_{G(fold)(T)}$ and $\beta_{G(unfold)(T)}$ are equivalent to the Brønsted exponents alpha and beta, respectively, and are interpreted to be a measure of the structural similarity of the transition state to either the reactants or products. If the introduction of a systematic perturbation (often a change in structure *via* addition or removal of a substituent, pH, solvent etc.) generates a reaction-series, and if for this reaction series it is found that alpha is close to zero (or beta close to unity), then it implies that the structure of the transition state is very similar to that of the reactant. Conversely, if alpha is close to unity (or beta is almost zero), it implies that the transition state is structurally similar to the product. Although the Brønsted exponents in many cases may be invariant with the degree of perturbation (i.e., a constant slope leading to linear free energy relationships),[36,37] this is not necessarily true, especially if the degree of perturbation is substantial (Figure 3 in Cohen and Marcus, 1968; Figure 1 in Kresge, 1975).[7,38,39] Further, this seemingly straightforward and logical Hammond-postulate-based conversion of Brønsted exponents to similarity or dissimilarity of the structure of the transition states to either of the ground states nevertheless fails for those systems with Brønsted exponents greater than unity and less than zero.[17,40-43]

In summary, there are three important general conclusions that we may draw from comparison of solvent ($\beta_{T(T)}$), entropic ($\beta_{G(T)}$) and heat capacity ($\beta_{H(T)}$) RCs: (*i*) as long as $\Delta SASA_{D-N}$ is large, positive and temperature-invariant, and by logical extension, $\Delta C_{pD-N}$ and $m_{D-N}$ are positive and temperature-invariant, the position of the TSE along the various RCs is neither constant nor a simple linear function of temperature when investigated over a large temperature range; (*ii*) for a given temperature, the position of the TSE along the RC depends on the choice of the RC; and (*iii*) although the algebraic sum of $\beta_{T(fold)(T)}$ and $\beta_{T(unfold)(T)}$, $\beta_{H(fold)(T)}$ and $\beta_{H(unfold)(T)}$, and $\beta_{G(fold)(T)}$ and $\beta_{G(unfold)(T)}$ must be unity for a two-state system for any particular temperature, individually they can be positive, negative, or zero. This is strongly supported by the occurrence of anomalous Brønsted exponents ($\beta < 0$ and $\beta > 1$) in physical organic chemistry, and anomalous $\Phi$-values ($\Phi < 0$ and $\Phi > 1$) in protein folding.[44,45] What this means for the $\Phi$-value-based canonical interpretation of the structure of the protein folding TSEs (a variation of the Brønsted procedure introduced by Fersht and coworkers wherein $0 \leq \Phi \leq 1$)[46] and the classical interpretation of the position of the TSE in reactions of small molecules is beyond the scope of this article and will be addressed elsewhere.

## CONCLUDING REMARKS

A system of equations that relate the position of the TSE along the RC to the various state functions has been derived for a spontaneously-folding fixed two-state system at constant pressure and solvent conditions using a treatment that is analogous to that given by Marcus for electron transfer. These equations allow the position of the TSE, the rate constants, the Gibbs energies, enthalpies, entropies, and heat capacities of activation for the folding and the unfolding to be ascertained at an unprecedented range and resolution of temperature provided a single thermal denaturation curve, a chevron, and the calorimetrically determined value of $\Delta C_{pD-N}$ are available. Although these equations have been developed for two-state proteins, they can further be extended to multi-state systems if the conformers that comprise the intermediate reaction-states, akin to those in the ground states, also behave like linear elastic springs, i.e., the Gibbs energies of the reactants and products have a square-law dependence on a suitable reaction coordinate, which incidentally, need not necessarily be SASA. Obviously, when the force constants for the reactants and products are identical and the mean length of the RC is normalized to unity, the equations reduce to those developed by Marcus.

As emphasized in Paper-I, the approximation that the mean length of the RC, $\Delta C_{p\text{D-N}}$, and the force constants are invariant for a given solvent at constant pressure applies only when temperature is the perturbant and need not be true, as we will show in subsequent publications, for other kinds of perturbations such as the change in primary sequence, addition of co-solvents, a change in the pressure, pH, primary sequence, etc. Consequently, the equations that describe the behaviour of two-systems when subjected to these perturbations will be far more complex, invariably requiring the use of multivariate calculus.

## COMPETING FINANCIAL INTERESTS

The author declares no competing financial interests.

## COPYRIGHT INFORMATION

This article is protected by copyright and is being released under Creative Commons Attribution-NonCommercial-NoDerivatives 4.0 International (CC BY-NC-ND 4.0; http://creativecommons.org/licenses/by-nc-nd/4.0/) license.

## APPENDIX

### The first derivatives of $m_{\text{TS-D}(T)}$ and $m_{\text{TS-N}(T)}$ with respect to temperature

If we recall that the force constants and $m_{\text{D-N}}$ are temperature-invariant, the first derivative of $m_{\text{TS-D}(T)}$ (Eq. (1)) with respect to temperature is given by

$$\frac{\partial m_{\text{TS-D}(T)}}{\partial T} = \frac{\partial}{\partial T}\left(\frac{\omega m_{\text{D-N}} - \sqrt{\varphi}}{(\omega - \alpha)}\right) = -\frac{1}{(\omega - \alpha)}\frac{\partial \sqrt{\varphi}}{\partial T} \tag{A1}$$

where $\varphi \equiv \lambda\omega + \Delta G_{\text{D-N}(T)}(\omega - \alpha)$. Now since $\sqrt{\varphi}$ is a composite function we use the chain rule to get

$$\frac{\partial \sqrt{\varphi}}{\partial T} = \frac{1}{2\sqrt{\varphi}}\frac{\partial \varphi}{\partial T} \tag{A2}$$

$$\frac{\partial \varphi}{\partial T} = \frac{\partial}{\partial T}\big(\lambda\omega + \Delta G_{\text{D-N}(T)}(\omega - \alpha)\big) = (\omega - \alpha)\frac{\partial \Delta G_{\text{D-N}(T)}}{\partial T} \tag{A3}$$

Substituting Eq. (A3) in (A2) gives

$$\frac{\partial \sqrt{\varphi}}{\partial T} = \frac{(\omega - \alpha)}{2\sqrt{\varphi}} \frac{\partial \Delta G_{\text{D-N}(T)}}{\partial T} \tag{A4}$$

Substituting the fundamental relationship $\partial \Delta G_{\text{D-N}(T)}/\partial T = -\Delta S_{\text{D-N}(T)}$ in (A4) gives

$$\frac{\partial \sqrt{\varphi}}{\partial T} = -\frac{(\omega - \alpha)\Delta S_{\text{D-N}(T)}}{2\sqrt{\varphi}} \tag{A5}$$

Substituting Eq. (A5) and (A1) yields

$$\frac{\partial m_{\text{TS-D}(T)}}{\partial T} = \frac{\Delta S_{\text{D-N}(T)}}{2\sqrt{\varphi}} \tag{A6}$$

The temperature-dependence of $\Delta S_{\text{D-N}(T)}$ is given by[4]

$$\Delta S_{\text{D-N}(T)} = \Delta S_{\text{D-N}(T_m)} + \int_{T_m}^{T} \frac{\Delta C_{p\text{D-N}(T)}}{T} dT = \Delta S_{\text{D-N}(T_m)} + \Delta C_{p\text{D-N}} \ln\left(\frac{T}{T_m}\right) \tag{A7}$$

where $\Delta S_{\text{D-N}(T)}$ and $\Delta S_{\text{D-N}(T_m)}$ denote the equilibrium entropies of unfolding, respectively, at any given temperature, and at the midpoint of thermal denaturation ($T_m$), respectively, for a given two-state folder under defined solvent conditions. The temperature-invariant and the temperature-dependent difference in heat capacity between the DSE and NSE is denoted by $\Delta C_{p\text{D-N}}$ and $\Delta C_{p\text{D-N}(T)}$, respectively. If $T_S$ (the temperature at which $\Delta S_{\text{D-N}(T)} = 0$) is used as the reference temperature, Eq. (A7) reduces to

$$\Delta S_{\text{D-N}(T)} = \Delta C_{p\text{D-N}} \ln\left(\frac{T}{T_S}\right) \tag{A8}$$

Substituting Eq. (A8) in (A6) gives the final form

$$\frac{\partial m_{\text{TS-D}(T)}}{\partial T} = \frac{\Delta C_{p\text{D-N}}}{2\sqrt{\varphi}} \ln\left(\frac{T}{T_S}\right) \begin{cases} < 0 \text{ for } T < T_S \\ = 0 \text{ for } T = T_S \\ > 0 \text{ for } T > T_S \end{cases} \tag{A9}$$

Because, $\beta_{\text{T(fold)}(T)} = m_{\text{TS-D}(T)}/m_{\text{D-N}}$ we also have

$$\frac{\partial \beta_{\text{T(fold)}(T)}}{\partial T} = \frac{1}{m_{\text{D-N}}} \frac{\partial m_{\text{TS-D}(T)}}{\partial T} = \frac{\Delta C_{p\text{D-N}}}{2m_{\text{D-N}}\sqrt{\varphi}} \ln\left(\frac{T}{T_S}\right) \tag{A10}$$

Since $\partial m_{\text{TS-D}(T)}/\partial T$ and $\partial \beta_{\text{T(fold)}(T)}/\partial T$ are physically undefined for $\varphi < 0$, their algebraic sign at any given temperature is determined by the $\ln(T/T_S)$ term. This leads to three scenarios: (*i*) for $T < T_S$ we have $\partial m_{\text{TS-D}(T)}/\partial T < 0$ and $\partial \beta_{\text{T(fold)}(T)}/\partial T < 0$; (*ii*) for $T > T_S$ we have $\partial m_{\text{TS-D}(T)}/\partial T > 0$ and $\partial \beta_{\text{T(fold)}(T)}/\partial T > 0$; and (*iii*) for $T = T_S$ we have $\partial m_{\text{TS-D}(T)}/\partial T = 0$ and $\partial \beta_{\text{T(fold)}(T)}/\partial T = 0$.

Since $m_{\text{TS-N}(T)} = (m_{\text{D-N}} - m_{\text{TS-D}(T)})$ and $\beta_{\text{T(unfold)}(T)} = m_{\text{TS-N}(T)}/m_{\text{D-N}}$, we have

$$\frac{\partial m_{\text{TS-N}(T)}}{\partial T} = -\frac{\partial m_{\text{TS-D}(T)}}{\partial T} = -\frac{\Delta S_{\text{D-N}(T)}}{2\sqrt{\varphi}} = \frac{\Delta C_{p\text{D-N}}}{2\sqrt{\varphi}} \ln\left(\frac{T_S}{T}\right) \begin{cases} > 0 \text{ for } T < T_S \\ = 0 \text{ for } T = T_S \\ < 0 \text{ for } T > T_S \end{cases} \quad (A11)$$

$$\frac{\partial \beta_{\text{T(unfold)}(T)}}{\partial T} = \frac{1}{m_{\text{D-N}}} \frac{\partial m_{\text{TS-N}(T)}}{\partial T} = \frac{\Delta C_{p\text{D-N}}}{2 m_{\text{D-N}} \sqrt{\varphi}} \ln\left(\frac{T_S}{T}\right) \quad (A12)$$

Once again, if we use the same argument as above, we end up with three scenarios: (*i*) for $T < T_S$ we have $\partial m_{\text{TS-N}(T)}/\partial T > 0$ and $\partial \beta_{\text{T(unfold)}(T)}/\partial T > 0$; (*ii*) for $T > T_S$ we have $\partial m_{\text{TS-N}(T)}/\partial T < 0$ and $\partial \beta_{\text{T(unfold)}(T)}/\partial T < 0$; and (*iii*) for $T = T_S$ we have $\partial m_{\text{TS-N}(T)}/\partial T = 0$ and $\partial \beta_{\text{T(unfold)}(T)}/\partial T = 0$.

## Expressions for activation entropies and enthalpies in terms of $\beta_{\text{T(fold)}(T)}$ and $\beta_{\text{T(unfold)}(T)}$

The activation entropy for folding in terms of $\beta_{\text{T(fold)}(T)}$ (also a composite function) may be readily derived by taking the first derivative of the second equality in Eq. (3) with respect to temperature. Therefore, we can write

$$\Delta S_{\text{TS-D}(T)} = -\frac{\partial}{\partial T}\left(\lambda \beta_{\text{T(fold)}(T)}^2\right) = -\lambda \frac{\partial \beta_{\text{T(fold)}(T)}^2}{\partial T} = -2\lambda \beta_{\text{T(fold)}(T)} \frac{\partial \beta_{\text{T(fold)}(T)}}{\partial T} \quad (A13)$$

Substituting Eq. (A10) in (A13) gives

$$\Delta S_{\text{TS-D}(T)} = \frac{\lambda \beta_{\text{T(fold)}(T)} \Delta C_{p\text{D-N}}}{m_{\text{D-N}} \sqrt{\varphi}} \ln\left(\frac{T_S}{T}\right) \quad (A14)$$

Similarly, the activation entropy for unfolding in terms of $\beta_{T(unfold)(T)}$ (also a composite function) may be readily derived by taking the first derivative of the second equality in Eq. (4) with respect to temperature. Therefore, we can write

$$\Delta S_{TS\text{-}N(T)} = -\frac{\partial}{\partial T}\left(\frac{\omega}{\alpha}\lambda\beta^2_{T(unfold)(T)}\right) = -\frac{\lambda\omega}{\alpha}\frac{\partial\beta^2_{T(unfold)(T)}}{\partial T} = -\frac{2\beta_{T(unfold)(T)}\lambda\omega}{\alpha}\frac{\partial\beta_{T(unfold)(T)}}{\partial T} \quad (A15)$$

Substituting Eq. (A12) in (A15) gives

$$\Delta S_{TS\text{-}N(T)} = \frac{\lambda\omega\beta_{T(unfold)(T)}\Delta C_{pD\text{-}N}}{\alpha m_{D\text{-}N}\sqrt{\varphi}}\ln\left(\frac{T}{T_S}\right) = \frac{\omega m_{D\text{-}N}\beta_{T(unfold)(T)}\Delta C_{pD\text{-}N}}{\sqrt{\varphi}}\ln\left(\frac{T}{T_S}\right) \quad (A16)$$

Substituting Eqs. (3) and (A14) in $\Delta H_{TS\text{-}D(T)} = \Delta G_{TS\text{-}D(T)} + T\Delta S_{TS\text{-}D(T)}$ and recasting gives

$$\Delta H_{TS\text{-}D(T)} = \lambda\beta^2_{T(fold)(T)}\left(1 + \frac{T\Delta C_{pD\text{-}N}}{\beta_{T(fold)(T)}m_{D\text{-}N}\sqrt{\varphi}}\ln\left(\frac{T_S}{T}\right)\right) \quad (A17)$$

Substituting Eqs. (4) and (A16) in $\Delta H_{TS\text{-}N(T)} = \Delta G_{TS\text{-}N(T)} + T\Delta S_{TS\text{-}N(T)}$ and recasting gives

$$\Delta H_{TS\text{-}N(T)} = \frac{\omega}{\alpha}\lambda\beta^2_{T(unfold)(T)}\left(1 + \frac{T\Delta C_{pD\text{-}N}}{\beta_{T(unfold)(T)}m_{D\text{-}N}\sqrt{\varphi}}\ln\left(\frac{T}{T_S}\right)\right) \quad (A18)$$

The reader will note that there are many other ways of recasting expressions for the temperature-dependence of the relevant state functions. For example, since the force constants are related to the variance of the Gaussian distribution of the SASA of the conformers in the DSE and the NSE, all of these equations can also be recast in terms of $\sigma^2_{DSE(T)}$ and $\sigma^2_{NSE(T)}$ or in terms of the partition functions of the DSE and NSE (see Paper-I).

**Expressions for the activation entropies at $T_m$ or $T_c$**

At the midpoint of thermal ($T_m$) or cold denaturation ($T_c$), $\Delta G_{D\text{-}N(T)} = 0$. Therefore, Eqs. (1) and (2) become

$$m_{TS\text{-}D(T)}\Big|_{T=T_c, T_m} = \frac{m_{D\text{-}N}\left(\omega - \sqrt{\alpha\omega}\right)}{(\omega - \alpha)} \Rightarrow \beta_{T(fold)(T)}\Big|_{T=T_c, T_m} = \frac{\omega - \sqrt{\alpha\omega}}{\omega - \alpha} \quad (A19)$$

$$m_{\text{TS-N}(T)}\Big|_{T=T_c,T_m} = \frac{m_{\text{D-N}}\left(\sqrt{\alpha\omega}-\alpha\right)}{(\omega-\alpha)} \Rightarrow \beta_{\text{T(unfold)}(T)}\Big|_{T=T_c,T_m} = \frac{\sqrt{\alpha\omega}-\alpha}{\omega-\alpha} \tag{A20}$$

Substituting Eq. (A19) and $\sqrt{\varphi_{(T)}}\Big|_{T=T_c,T_m} = \sqrt{\lambda\omega} = m_{\text{D-N}}\sqrt{\alpha\omega}$ in (9) and simplifying gives

$$\Delta S_{\text{TS-D}(T)}\Big|_{T=T_c,T_m} = \frac{\sqrt{\alpha\omega}-\alpha}{(\omega-\alpha)}\Delta C_{p\text{D-N}}\ln\left(\frac{T_S}{T}\right)\Big|_{T=T_c,T_m} \begin{cases} >0 \text{ for } T=T_c \\ <0 \text{ for } T=T_m \end{cases} \tag{A21}$$

Similarly, substituting Eq. (A20) in (11) and simplifying gives:

$$\Delta S_{\text{TS-N}(T)}\Big|_{T=T_c,T_m} = \frac{\omega-\sqrt{\alpha\omega}}{(\omega-\alpha)}\Delta C_{p\text{D-N}}\ln\left(\frac{T}{T_S}\right)\Big|_{T=T_c,T_m} \begin{cases} <0 \text{ for } T=T_c \\ >0 \text{ for } T=T_m \end{cases} \tag{A22}$$

Since the mid-point of thermal denaturation can be experimentally determined with fairly good accuracy and precision, and is a defining constant for any given two-state folder when pressure and solvent conditions are defined, Eqs. (A21) and (A22) allow $\Delta S_{\text{TS-D}(T)}$ and $\Delta S_{\text{TS-N}(T)}$ to be quickly calculated for this particular temperature if the force constants, $\Delta C_{p\text{D-N}}$, $T_S$, and $T_m$ are known. Because the parameters in these equations are all temperature-invariant, these will be defining constants for two-state-folding primary sequence when pressure and solvent are defined.

**The relative positions of the reference temperatures**

Protein stability curves are usually described from the perspective of four equilibrium reference temperatures which are: (*i*) the cold and heat denaturation temperatures, $T_c$ and $T_m$, respectively, at which $\Delta G_{\text{D-N}(T)}= 0$; (*ii*) $T_H$, the temperature at which $\Delta H_{\text{D-N}(T)}= 0$; and (*iii*) $T_S$, the temperature at which $\Delta S_{\text{D-N}(T)}= 0$. The relative position of these four references temperatures along the temperature axis is given by: $T_c < T_H < T_S < T_m$.[4,24] From the temperature-dependence of the rate constants we have two additional reference temperatures which are: (*i*) $T_{H(\text{TS-N})}$, the temperature at which $\Delta H_{\text{TS-N}(T)}= 0$; and (*ii*) $T_{H(\text{TS-D})}$, the temperature at which $\Delta H_{\text{TS-D}(T)}= 0$. In the discussion on the temperature-dependence of the algebraic sign of $\Delta H_{\text{TS-D}(T)}$, we have shown that $T_S < T_{H(\text{TS-D})}$; and in the discussion on the temperature-dependence of the algebraic sign of $\Delta H_{\text{TS-N}(T)}$, we have shown that $T_{H(\text{TS-N})} < T_S$. This leads to the first logical conclusion: $T_{H(\text{TS-N})} < T_S < T_{H(\text{TS-D})}$. Because $\Delta H_{\text{D-N}(T)} = \Delta H_{\text{TS-N}(T)} - \Delta H_{\text{TS-D}(T)}$ for a two-state system, for $\Delta H_{\text{D-N}(T)}$ to be zero at $T_H$, $\Delta H_{\text{TS-N}(T)}$ and $\Delta H_{\text{TS-D}(T)}$ must be identical at $T_H$, or $\Delta H_{\text{TS-N}(T)}$ and $\Delta H_{\text{TS-D}(T)}$ functions must intersect. Because $\Delta H_{\text{TS-D}(T)} > 0$ for $T < T_{H(\text{TS-D})}$ and since $T_H < T_S < T_{H(\text{TS-D})}$, $\Delta H_{\text{TS-N}(T)}$ must also be positive for the intersection to occur. Since $\Delta H_{\text{TS-N}(T)} > 0$ only when $T > T_{H(\text{TS-N})}$, the logical conclusion is that $T_H > T_{H(\text{TS-N})}$. In other words, $T_H$ is located between $T_{H(\text{TS-N})}$ and $T_S$. The mathematical formalism is as follows: The temperature-dependence of $\Delta H_{\text{D-N}(T)}$ is given by[4]

$$\Delta H_{\text{D-N}(T)} = \Delta H_{\text{D-N}(T_m)} + \int_{T_m}^{T} \Delta C_{p\text{D-N}(T)}\, dT = \Delta H_{\text{D-N}(T_m)} + \Delta C_{p\text{D-N}}(T - T_m) \tag{A23}$$

where $\Delta H_{\text{D-N}(T)}$ and $\Delta H_{\text{D-N}(T_m)}$ denote the equilibrium enthalpies of unfolding, respectively, at any given temperature, and at the midpoint of thermal denaturation ($T_m$), respectively, for a given two-state folder under constant pressure and solvent conditions. When $T_H$ is used as the reference temperature, Eq. (A23) becomes

$$\Delta H_{\text{D-N}(T)} = \Delta C_{p\text{D-N}}(T - T_H) \tag{A24}$$

Thus, when $T = T_{H(\text{TS-N})}$, Eq. (A24) becomes

$$\Delta H_{\text{D-N}(T)}\big|_{T=T_{H(\text{TS-N})}} = \Delta C_{p\text{D-N}}\left(T_{H(\text{TS-N})} - T_H\right) \tag{A25}$$

For a two-state system we have $\Delta H_{\text{D-N}(T)} = \Delta H_{\text{TS-N}(T)} - \Delta H_{\text{TS-D}(T)}$. Because $\Delta H_{\text{TS-N}(T)} = 0$ at $T_{H(\text{TS-N})}$, Eq. (A25) can be rearranged to give

$$\Delta H_{\text{TS-D}(T)}\big|_{T=T_{H(\text{TS-N})}} = \Delta C_{p\text{D-N}}\left(T_H - T_{H(\text{TS-N})}\right) \tag{A26}$$

Because $\Delta H_{\text{TS-D}(T)} > 0$ for $T < T_{H(\text{TS-D})}$ and $T_{H(\text{TS-N})} < T_S < T_{H(\text{TS-D})}$, the left-hand-side (LHS) in Eq. (A26) is positive. Now since $\Delta C_{p\text{D-N}}$ is positive, the condition that the LHS must be positive is satisfied if and only if $T_H > T_{H(\text{TS-N})}$. This leads to the relationship: $T_{H(\text{TS-N})} < T_H < T_S < T_{H(\text{TS-D})}$. This relationship must hold for the reversible thermal transitions of all two-state folders that conform to the postulates laid out in Paper-I.

# The second derivatives of $m_{\text{TS-D}(T)}$ and $m_{\text{TS-N}(T)}$ with respect to temperature

Differentiating Eq. (A6) with respect to temperature gives

$$\frac{\partial^2 m_{\text{TS-D}(T)}}{\partial T^2} = \frac{\partial}{\partial T}\left(\frac{\Delta S_{\text{D-N}(T)}}{2\sqrt{\varphi}}\right) = \frac{1}{2}\frac{\partial}{\partial T}\left(\frac{\Delta S_{\text{D-N}(T)}}{\sqrt{\varphi}}\right) \quad (A27)$$

Using the quotient rule we get

$$\frac{\partial^2 m_{\text{TS-D}(T)}}{\partial T^2} = \frac{1}{2}\left(\frac{\sqrt{\varphi}\frac{\partial \Delta S_{\text{D-N}(T)}}{\partial T} - \Delta S_{\text{D-N}(T)}\frac{\partial \sqrt{\varphi}}{\partial T}}{\left(\sqrt{\varphi}\right)^2}\right) \quad (A28)$$

Substituting $\partial \Delta S_{\text{D-N}(T)}/\partial T = \Delta C_{p\text{D-N}}/T$ in Eq. (A28) gives

$$\frac{\partial^2 m_{\text{TS-D}(T)}}{\partial T^2} = \frac{1}{2\varphi}\left(\sqrt{\varphi}\frac{\Delta C_{p\text{D-N}}}{T} - \Delta S_{\text{D-N}(T)}\frac{\partial \sqrt{\varphi}}{\partial T}\right) \quad (A29)$$

Substituting Eq. (A5) in (A29) gives

$$\frac{\partial^2 m_{\text{TS-D}(T)}}{\partial T^2} = \frac{1}{2\varphi}\left(\sqrt{\varphi}\frac{\Delta C_{p\text{D-N}}}{T} + \frac{(\omega-\alpha)\left(\Delta S_{\text{D-N}(T)}\right)^2}{2\sqrt{\varphi}}\right) \quad (A30)$$

Simplifying Eq. (A30) yields the final form

$$\frac{\partial^2 m_{\text{TS-D}(T)}}{\partial T^2} = \frac{1}{4T\varphi\sqrt{\varphi}}\left[2\varphi\Delta C_{p\text{D-N}} + T\left(\Delta S_{\text{D-N}(T)}\right)^2(\omega-\alpha)\right] \quad (A31)$$

Similarly, we may show that

$$\frac{\partial^2 m_{\text{TS-N}(T)}}{\partial T^2} = -\frac{\partial^2 m_{\text{TS-D}(T)}}{\partial T^2} = -\frac{1}{4T\varphi\sqrt{\varphi}}\left[2\varphi\Delta C_{p\text{D-N}} + T\left(\Delta S_{\text{D-N}(T)}\right)^2(\omega-\alpha)\right] \quad (A32)$$

Eqs. (A31) and (A32) will be useful in deriving expressions for the heat capacities of activation for folding and unfolding as shown below.

## Expressions for the heat capacities of activation for folding and unfolding

The expressions for the heat capacities of activation for folding and unfolding may be obtained by differentiating the expressions for the activation enthalpies or entropies with respect to temperature. Differentiating Eq. (8) yields:

$$\frac{\partial \Delta S_{\text{TS-D}(T)}}{\partial T} = \frac{\Delta C_{p\text{TS-D}(T)}}{T} = -2\alpha \frac{\partial}{\partial T}\left( m_{\text{TS-D}(T)} \frac{\partial m_{\text{TS-D}(T)}}{\partial T}\right) \tag{A33}$$

Using the product rule we get

$$\frac{\Delta C_{p\text{TS-D}(T)}}{T} = -2\alpha \left( m_{\text{TS-D}(T)} \frac{\partial^2 m_{\text{TS-D}(T)}}{\partial T^2} + \left(\frac{\partial m_{\text{TS-D}(T)}}{\partial T}\right)^2 \right) \tag{A34}$$

From Eq. (A6) we have

$$\left(\frac{\partial m_{\text{TS-D}(T)}}{\partial T}\right)^2 = \left(\frac{\Delta S_{\text{D-N}(T)}}{2\sqrt{\varphi}}\right)^2 = \frac{\left(\Delta S_{\text{D-N}(T)}\right)^2}{4\varphi} \tag{A35}$$

Substituting Eqs. (A31) and (A35) in (A34) and simplifying gives

$$\frac{\Delta C_{p\text{TS-D}(T)}}{\cancel{T}} = -\frac{\alpha}{2\cancel{T}\varphi\sqrt{\varphi}}\left[\left(m_{\text{TS-D}(T)}\left(2\varphi \Delta C_{p\text{D-N}} + T\left(\Delta S_{\text{D-N}(T)}\right)^2 (\omega-\alpha)\right)\right) + T\sqrt{\varphi}\left(\Delta S_{\text{D-N}(T)}\right)^2 \right] \tag{A36}$$

$$\Delta C_{p\text{TS-D}(T)} = -\frac{\alpha}{2\varphi\sqrt{\varphi}}\left[ m_{\text{TS-D}(T)} 2\varphi \Delta C_{p\text{D-N}} + \left(m_{\text{TS-D}(T)} T\left(\Delta S_{\text{D-N}(T)}\right)^2 (\omega-\alpha)\right) + T\sqrt{\varphi}\left(\Delta S_{\text{D-N}(T)}\right)^2\right]$$
$$= -\frac{\alpha}{2\varphi\sqrt{\varphi}}\left[\left(m_{\text{TS-D}(T)} 2\varphi \Delta C_{p\text{D-N}}\right) + T\left(\Delta S_{\text{D-N}(T)}\right)^2 \left(m_{\text{TS-D}(T)}(\omega-\alpha) + \sqrt{\varphi}\right)\right] \tag{A37}$$

Substituting $m_{\text{TS-D}(T)}(\omega-\alpha) = \omega m_{\text{D-N}} - \sqrt{\varphi}$ (see Eq. (1)) in Eq. (A37) gives

$$\Delta C_{p\text{TS-D}(T)} = -\frac{\alpha}{2\varphi\sqrt{\varphi}}\left[ m_{\text{TS-D}(T)} 2\varphi \Delta C_{p\text{D-N}} + T\left(\Delta S_{\text{D-N}(T)}\right)^2 \left(\omega m_{\text{D-N}} - \cancel{\sqrt{\varphi}} + \cancel{\sqrt{\varphi}}\right)\right]$$
$$= -\frac{\alpha}{2\varphi\sqrt{\varphi}}\left[ m_{\text{TS-D}(T)} 2\varphi \Delta C_{p\text{D-N}} + \omega m_{\text{D-N}} T\left(\Delta S_{\text{D-N}(T)}\right)^2\right] \tag{A38}$$

$$\Rightarrow \Delta C_{p\text{D-TS}(T)} = \frac{\alpha}{2\varphi\sqrt{\varphi}}\left[m_{\text{TS-D}(T)}2\varphi\Delta C_{p\text{D-N}} + \omega m_{\text{D-N}}T\left(\Delta S_{\text{D-N}(T)}\right)^2\right] \quad (A39)$$

Similarly, differentiating Eq. (10) with respect to temperature gives

$$\frac{\partial \Delta S_{\text{TS-N}(T)}}{\partial T} = \frac{\Delta C_{p\text{TS-N}(T)}}{T} = -2\omega\frac{\partial}{\partial T}\left(m_{\text{TS-N}(T)}\frac{\partial m_{\text{TS-N}(T)}}{\partial T}\right) \quad (A40)$$

Using the product rule gives

$$\frac{\Delta C_{p\text{TS-N}(T)}}{T} = -2\omega\left(m_{\text{TS-N}(T)}\frac{\partial^2 m_{\text{TS-N}(T)}}{\partial T^2} + \left(\frac{\partial m_{\text{TS-N}(T)}}{\partial T}\right)^2\right) \quad (A41)$$

Recasting Eq. (A41) in terms of (A11) and (A32) gives

$$\frac{\Delta C_{p\text{TS-N}(T)}}{T} = 2\omega\left(m_{\text{TS-N}(T)}\frac{\partial^2 m_{\text{TS-D}(T)}}{\partial T^2} - \left(-\frac{\partial m_{\text{TS-D}(T)}}{\partial T}\right)^2\right) \quad (A42)$$

Substituting Eqs. (A31) and (A35) in (A42) and simplifying gives

$$\Delta C_{p\text{TS-N}(T)} = \frac{\omega}{2\varphi\sqrt{\varphi}}\left[\left(m_{\text{TS-N}(T)}2\varphi\Delta C_{p\text{D-N}}\right) + T\left(\Delta S_{\text{D-N}(T)}\right)^2\left(m_{\text{TS-N}(T)}(\omega-\alpha)-\sqrt{\varphi}\right)\right] \quad (A43)$$

Substituting $m_{\text{TS-N}(T)}(\omega-\alpha) = \sqrt{\varphi} - \alpha m_{\text{D-N}}$ (see Eq. (2)) in Eq. (A43) and simplifying yields the final form

$$\Delta C_{p\text{TS-N}(T)} = \frac{\omega}{2\varphi\sqrt{\varphi}}\left[m_{\text{TS-N}(T)}2\varphi\Delta C_{p\text{D-N}} - \alpha m_{\text{D-N}}T\left(\Delta S_{\text{D-N}(T)}\right)^2\right] \quad (A44)$$

Although not shown, the algebraic sum of Eqs. (A39) and (A44) will be equal to $\Delta C_{p\text{D-N}}$. Furthermore, both these equations can also be derived from differentiating activation enthalpies with respect to temperature.

## Expressions for $\Delta C_{p\text{D-TS}(T)}$ and $\Delta C_{p\text{TS-N}(T)}$ when $T = T_{S(\alpha)}$ and $T = T_{S(\omega)}$

Since $m_{\text{TS-N}(T)} = 0$, $\sqrt{\varphi} = \alpha m_{\text{D-N}}$ and $\varphi = (\alpha m_{\text{D-N}})^2$ at $T_{S(\alpha)}$ and $T_{S(\omega)}$, Eq. (A44) reduces to

$$\Delta C_{p\text{TS-N}(T)}\Big|_{T=T_{S(\alpha)},T_{S(\omega)}} = -\frac{\omega\alpha m_{\text{D-N}}T\left(\Delta S_{\text{D-N}(T)}\right)^2}{2\varphi\sqrt{\varphi}}\Bigg|_{T=T_{S(\alpha)},T_{S(\omega)}} = -\frac{\omega T}{2}\left(\frac{\Delta S_{\text{D-N}(T)}}{\alpha m_{\text{D-N}}}\right)^2\Bigg|_{T=T_{S(\alpha)},T_{S(\omega)}} \quad (A45)$$

Further, since $\Delta C_{p\text{D-N}} = \Delta C_{p\text{D-TS}(T)} + \Delta C_{p\text{TS-N}(T)}$ we have

$$\Delta C_{p\text{D-TS}(T)}\Big|_{T=T_{S(\alpha)},T_{S(\omega)}} = \Delta C_{p\text{D-N}} + \frac{\omega T}{2}\left(\frac{\Delta S_{\text{D-N}(T)}}{\alpha m_{\text{D-N}}}\right)^2\Bigg|_{T=T_{S(\alpha)},T_{S(\omega)}} > \Delta C_{p\text{D-N}} \quad (A46)$$

Because $\Delta C_{p\text{TS-N}(T)} < 0$ at $T_{S(\alpha)}$ and $T_{S(\omega)}$, and the lone extremum of $\Delta C_{p\text{TS-N}(T)}$ (which is algebraically positive and a maximum) occurs at $T_S$, it implies that there will be two unique temperatures at which $\Delta C_{p\text{TS-N}(T)} = 0$, one in the low temperature ($T_{C_p\text{TS-N}(\alpha)}$) such that $T_{S(\alpha)} < T_{C_p\text{TS-N}(\alpha)} < T_S$, and the other in the high temperature regime ($T_{C_p\text{TS-N}(\omega)}$) such that $T_S < T_{C_p\text{TS-N}(\omega)} < T_{S(\omega)}$. Thus, at the these two unique temperatures $T_{C_p\text{TS-N}(\alpha)}$ and $T_{C_p\text{TS-N}(\omega)}$, we have $\Delta C_{p\text{D-TS}(T)} = \Delta C_{p\text{D-N}} \Rightarrow \beta_{\text{H(fold)}(T)} = 1$ and $\beta_{\text{H(unfold)}(T)} = 0$; and for the temperature regimes $T_\alpha \leq T < T_{C_p\text{TS-N}(\alpha)}$ and $T_{C_p\text{TS-N}(\omega)} < T \leq T_\omega$, we have $\Delta C_{p\text{D-TS}(T)} > \Delta C_{p\text{D-N}} \Rightarrow \beta_{\text{H(fold)}(T)} > 1$, and $\Delta C_{p\text{TS-N}(T)} < 0 \Rightarrow \beta_{\text{H(unfold)}(T)} < 0$. The prediction that $\Delta C_{p\text{TS-N}(T)}$ must approach zero at some high temperature is readily apparent from data on chymotrypsin inhibitor-2: Despite the temperature-range not being substantial (320 to 340 K), and the data points that define the $\Delta H_{\text{TS-N}(T)}$ function being sparse (7 in total), it is apparent even from a casual inspection that it is clearly non-linear with temperature (Figure 5B in Tan et al., 1996).[47] Although Fersht and co-workers have fitted the data to linear function and reached the natural conclusion that the difference in heat capacity between the TSE and the NSE is temperature-invariant, they nevertheless explicitly mention that if the non-linearity of $\Delta H_{\text{TS-N}(T)}$ were given due consideration, and the data are fit to an empirical-quadratic instead of a linear function, $\Delta C_{p\text{TS-N}(T)}$ indeed becomes temperature-dependent and is predicted to approach zero at $\approx 360$ K (see text in page 382 in Tan et al., 1996).[47]

Now, since $\Delta C_{p\text{TS-N}(T)} > 0$ and a maximum, and $\Delta C_{p\text{D-TS}(T)} > 0$ and a minimum at $T_S$, and decrease and increase, respectively, with any deviation in temperature from $T_S$, and since $\Delta C_{p\text{TS-N}(T)}$ becomes zero at $T_{C_p\text{TS-N}(\alpha)}$ and $T_{C_p\text{TS-N}(\omega)}$, the obvious mathematical consequence is that $\Delta C_{p\text{D-TS}(T)}$ and $\Delta C_{p\text{TS-N}(T)}$ functions must intersect at two unique temperatures. Because at the point of intersection we have the relationship: $\Delta C_{p\text{D-TS}(T)} = \Delta C_{p\text{TS-N}(T)} = \Delta C_{p\text{D-N}}/2$, a consequence is that $\Delta C_{p\text{TS-N}(T)}$ must be positive at the point of intersection, with the low-temperature intersection occurring between $T_{C_p\text{TS-N}(\alpha)}$ and $T_S$, and the high-temperature intersection between $T_S$ and $T_{C_p\text{TS-N}(\omega)}$. An equivalent interpretation is that the absolute heat capacity of the TSE is exactly half the algebraic sum of the absolute heat capacities of the DSE and the NSE at the temperatures where $\Delta C_{p\text{D-TS}(T)}$ and $\Delta C_{p\text{TS-N}(T)}$ intersect.

## Expressions for $\beta_{G(\text{fold})(T)}$ and $\beta_{G(\text{unfold})(T)}$ when $T = T_c$ and $T = T_m$

Substituting Eqs. (A8), (A21) and (A22) in (41) and (43), and simplifying gives (detailed steps not shown)

$$\beta_{G(\text{fold})(T)}\Big|_{T=T_c,T_m} = \frac{\sqrt{\alpha\omega} - \alpha}{\omega - \alpha} \equiv \beta_{T(\text{unfold})(T)}\Big|_{T=T_c,T_m} \tag{A47}$$

$$\beta_{G(\text{unfold})(T)}\Big|_{T=T_c,T_m} = \frac{\omega - \sqrt{\alpha\omega}}{\omega - \alpha} \equiv \beta_{T(\text{fold})(T)}\Big|_{T=T_c,T_m} \tag{A48}$$

Simply put, at the midpoint of cold or heat denaturation, the position of the TSE relative to the DSE along the normalized entropic RC is identical to the position of the TSE relative to the NSE along the normalized SASA-RC. Similarly, the position of the TSE relative to the NSE along the normalized entropic RC is identical to the position of the TSE relative to the DSE along the normalized SASA-RC. Dividing Eq. (A47) by (A48) gives

$$\frac{\beta_{G(\text{fold})(T)}}{\beta_{G(\text{unfold})(T)}}\Bigg|_{T=T_c,T_m} = \frac{\beta_{T(\text{unfold})(T)}}{\beta_{T(\text{fold})(T)}}\Bigg|_{T=T_c,T_m} \Rightarrow \frac{-\Delta S_{\text{TS-D}(T)}}{\Delta S_{\text{TS-N}(T)}}\Bigg|_{T=T_c,T_m} = \frac{m_{\text{TS-N}(T)}}{m_{\text{TS-D}(T)}}\Bigg|_{T=T_c,T_m} \tag{A49}$$

This seemingly obvious relationship has far deeper physical meaning. Simplifying further and recasting gives

$$\frac{\Delta S_{\text{TS-N}(T)}}{\Delta S_{\text{D-TS}(T)}}\Bigg|_{T=T_c,T_m} = \frac{m_{\text{TS-D}(T)}}{m_{\text{TS-N}(T)}}\Bigg|_{T=T_c,T_m} = \sqrt{\frac{\omega}{\alpha}}\Bigg|_{T=T_c,T_m} = \sqrt{\frac{\sigma^2_{\text{DSE}(T)}}{\sigma^2_{\text{NSE}(T)}}}\Bigg|_{T=T_c,T_m} = \frac{\sigma_{\text{DSE}(T)}}{\sigma_{\text{NSE}(T)}}\Bigg|_{T=T_c,T_m} \tag{A50}$$

Thus, when the concentration of the DSE and the NSE are identical, the ratio of the slopes of the folding and unfolding arms of the chevron are a measure of the ratio of the change in entropies for the partial unfolding reactions $N \rightleftharpoons [TS]$ and $[TS] \rightleftharpoons D$, or the ratio of the standard deviations of the DSE and the NSE Gaussians along the SASA-RC.

**Integrated Gibbs-Helmholtz equation**

The temperature-dependence of $\Delta G_{\text{D-N}(T)}$ is given by the Gibbs-Helmholtz equation[4]

$$\Delta G_{\text{D-N}(T)} = \Delta H_{\text{D-N}(T_m)}\left(1 - \frac{T}{T_m}\right) + \Delta C_{p\text{D-N}}(T - T_m) + T\Delta C_{p\text{D-N}} \ln\left(\frac{T_m}{T}\right) \qquad (A51)$$

where $\Delta H_{\text{D-N}(T_m)}$ is the enthalpy of unfolding at the midpoint of thermal denaturation.

# Table 1: Glossary of reference temperatures

| Temperature | Remark |
|---|---|
| $T_\alpha$ | A two-state system is physically undefined for $T < T_\alpha$ |
| $T_{S(\alpha)}$ | $\Delta H_{\text{TS-N}(T)} = \Delta S_{\text{TS-N}(T)} = \Delta G_{\text{TS-N}(T)} = 0$, $k_{u(T)} = k^0$ |
| $T_{C_p\text{TS-N}(\alpha)}$ | $\Delta C_{p\text{TS-N}(T)} = 0$ |
| $T_c$ | Midpoint of cold denaturation, $\Delta G_{\text{D-N}(T)} = 0$, $k_{f(T)} = k_{u(T)}$ |
| $T_{H(\text{TS-N})}$ | $\Delta H_{\text{TS-N}(T)} = 0$, $k_{u(T)}$ is a minimum |
| $T_H$ | $\Delta H_{\text{TS-D}(T)} = \Delta H_{\text{TS-N}(T)}$, $\Delta H_{\text{D-N}(T)} = 0$, $\Delta H_{\text{TS-D}(T)} > 0$, $\Delta H_{\text{TS-N}(T)} > 0$, |
| $T_S$ | $\Delta S_{\text{TS-D}(T)} = \Delta S_{\text{TS-N}(T)} = \Delta S_{\text{D-N}(T)} = 0$, $\Delta G_{\text{D-N}(T)}$ is a maximum |
| $T_{H(\text{TS-D})}$ | $\Delta H_{\text{TS-D}(T)} = 0$, $k_{f(T)}$ is a maximum |
| $T_m$ | Midpoint of heat denaturation, $\Delta G_{\text{D-N}(T)} = 0$, $k_{f(T)} = k_{u(T)}$ |
| $T_{C_p\text{TS-N}(\omega)}$ | $\Delta C_{p\text{TS-N}(T)} = 0$ |
| $T_{S(\omega)}$ | $\Delta H_{\text{TS-N}(T)} = \Delta S_{\text{TS-N}(T)} = \Delta G_{\text{TS-N}(T)} = 0$, $k_{u(T)} = k^0$ |
| $T_\omega$ | A two-state system is physically undefined for $T > T_\omega$ |

# FIGURES

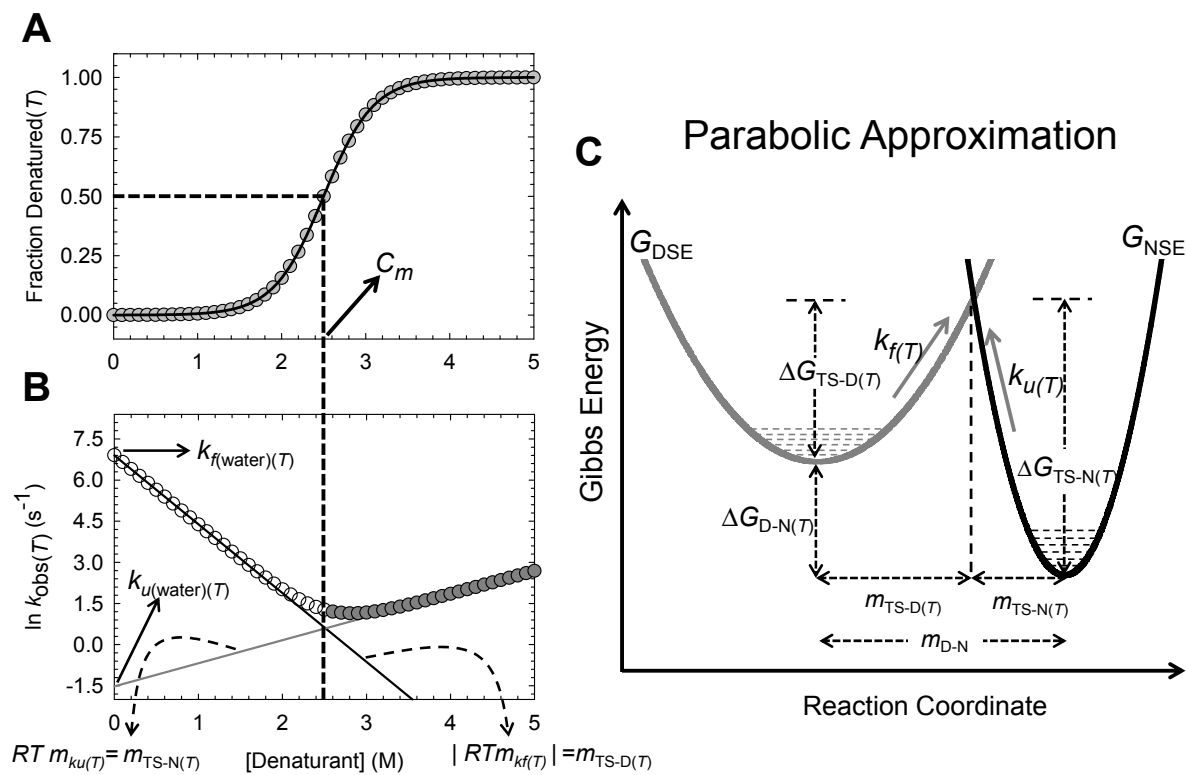

**Figure 1.**

**Standard equilibrium and kinetic parameters from chemical denaturation and their relationship to parabolic Gibbs energy curves.**

(A) An equilibrium chemical denaturation curve simulated using standard two-state equations for a hypothetical two-state protein with $\Delta G_{D-N(T)}$ = 5 kcal.mol$^{-1}$; $m_{D-N}$ = 2 kcal.mol$^{-1}$.M$^{-1}$; $C_m$ ≡ [Den$_{50\%}$] = 2.5 M and $T$ = 298.16 K.[48] The midpoint of chemical denaturation is given by [Den$_{50\%}$]. (B) A corresponding chevron simulated using the standard chevron-equation with $k_{f(water)(T)}$ = 1000 s$^{-1}$; $k_{u(water)(T)}$ = 0.216 s$^{-1}$; $m_{TS-D(T)}$ and $m_{TS-N(T)}$ are 1.5 and 0.5 kcal.mol$^{-1}$.M$^{-1}$, respectively, and $T$ = 298.16 K.[48] The denaturant-dependences of ln $k_{f(T)}$ and ln $k_{u(T)}$ are given by $m_{kf(T)}$ (solid black line) and $m_{ku(T)}$ (solid grey line), respectively. (C) Parabolic approximation for a two-state protein. The Gibbs barrier heights for folding and unfolding are given by $\Delta G_{TS-D(T)}$ and $\Delta G_{TS-N(T)}$, respectively. The mean length of the RC is given by $m_{D-N}$, the position of the TSE with respect to the DSE and the NSE along the RC are given by $m_{TS-D(T)}$ and $m_{TS-N(T)}$, respectively.